\documentclass[12pt,preprint]{aastex}

\shorttitle{Energy balances of the Dec. 13 2001 CME}
\shortauthors{Lee et al.}

\begin{document}

\title{Three Dimensional Structure and Energy Balance of a Coronal Mass Ejection}

\author{J.-Y. Lee\altaffilmark{1,2,3}, J. C. Raymond\altaffilmark{2}, Y.-K. Ko\altaffilmark{2}, and K.-S. Kim\altaffilmark{1}}

\altaffiltext{1}{Dept. of Astronomy and Space
Science, Kyung Hee University, Yongin, Gyeonggi, 446-701, Korea}
\altaffiltext{2}{Harvard-Smithsonian Center for Astrophysics,
Cambridge, MA 02138}
\altaffiltext{3}{NorthWest Research Associates, CoRA Division, Boulder, CO 80301}

\begin{abstract}
The Ultraviolet Coronagraph Spectrometer (UVCS) observed Doppler shifted material of a partial Halo Coronal Mass Ejection (CME) on December 13 2001. The observed ratio of [O V]/O V] is a reliable density diagnostic important for assessing the state of the plasma. Earlier UVCS observations of CMEs found evidence that the ejected plasma is heated long after the eruption. We have investigated the heating rates, which represent a significant fraction of the CME energy budget. The parameterized heating and radiative and adiabatic cooling have been used to evaluate the temperature evolution of the CME material with a time dependent ionization state model. The functional form of a flux rope model for interplanetary magnetic clouds was also used to parameterize the heating. We find that continuous heating is required to match the UVCS observations. To match
the O VI-bright knots, a higher heating rate is required such that
the heating energy is greater than the kinetic energy. The
temperatures for the knots bright in Ly$\alpha$ and C III emission indicate
that smaller heating rates are required for those regions. In the context of the flux rope model, about 75\% of the magnetic energy must go into heat in order to match the O VI observations. We derive tighter constraints on the heating than earlier analyses, and we show that thermal conduction with the Spitzer conductivity is not sufficient to account for the heating at large heights.

\end{abstract}


\section{Introduction}

The UVCS on board the {\it
Solar and Heliospheric Observatory} (SOHO) provides unique
spectroscopic diagnostics of CME material
through both Doppler shifts and line intensities \citep{ray02}. For CME studies, major questions are the three dimensional structure
and the energy budget of the ejected CME plasma. UVCS observations
contribute to the understanding of these two major problems. The temperature evolution and energy balance during the ejection of
CME have been studied by several authors \citep{ak01, ku96, em04,
vo00, su07}. Many studies of CME three dimensional structure
have been based on the two dimensional projection of the 3D structure on the plane of the sky and focused on the efforts to overcome that loss of information (see Burkepile et al. 2004 and the references in that paper). In this paper, we present the three dimensional structure, heating, and
energy balances for a CME observed on Dec. 13, 2001.

The kinetic properties of CMEs from measurements of white light
coronal observations have been determined, but suffer from the inaccuracies due to projection effects \citep{burkepile04}. Those make it
difficult to study of the three-dimensional structure. To overcome
the projection effect, several geometrical assumptions such as the cone model \citep{zh02,
zie04, yeh05}, ice-cream cone model \citep{fimu84, xu05}, and the time
difference between the appearances of the halo at two opposite
position angles \citep{mi03} have been applied to evaluate the CME
properties. The images reconstructed through polarization analysis indicate that CME is a bubble-like structure rather than an expanding loop
arcade structure \citep{cpc83, mo04, de06}. Howard et al. (2007) used SMEI data with the assumption of  purely radial expansion from the eruption site to estimate the position of the ejecta in three dimensions. The three dimensional
structure reconstructed with UV emission lines observed by UVCS shows
a ribbon-like structure in one event \citep{cia03} and a halo CME structure in the view from the solar west in another \citep{lee06}. The flux-rope model
\citep{kr06} compared statistically with CME observed by {\it
Solar Maximum Mission} (SMM) shows that the morphology of the CME
is hollow \citep{kr07}. The Solar Terrestrial Relations Observatory
(STEREO) was launched on 25th Oct. 2006, and two spacecraft
trailing and leading Earth will provide future studies of the
three dimensional structure of CME \citep{ka05}.

The line-of-sight velocity component observed by UVCS makes it possible to
estimate the angle with the plane of the sky. For evaluating the required heating for the UVCS observations, this angle is used to
estimate the radial velocities of the ejected plasma, and the column
density from Large Angle and Spectrometric Coronagraph (LASCO) measurements. The angles with the plane of the sky have been evaluated
for the 22 halo and partial halo CME cores observed by UVCS to
estimate projection effects \citep{cia06}.

The energy partitions for flare-CME events show that the CME has the dominant
component of the released energy, and that it contains a substantial
fraction of the available magnetic energy \citep{em05}. Studies of flux rope CME kinetic energies also show that internal magnetic energy is a
viable source of CME energy \citep{su07, vo00}. Earlier UVCS
observations of CME found evidence that the ejected plasma is heated
long after the eruption \citep{ak01, cia01}, and that the total heat going into the ejected plasma is comparable to the kinetic energy. 
Other studies based on the thermal energy evolution of ICMEs \citep{liu06}, emission in the EUV bands of EIT \citep{fiko02} and the ionization states measured in ICMEs \citep{ra07} also found that CME plasma is strongly heated even after it leaves the solar surface.

In this paper, we find the heating by a procedure similar to that of Akmal et al. (2001).
We first generate a large grid of models with different initial conditions and forms of the heating function. Then, for individual knots observed by UVCS, we select all the models that match the observed spectra, and the range of heat inputs for those models is the range of heating allowed.

We find that continuous heating is required to match the UVCS
observations. The temperature evolution shows a rapid decrease at lower heights and earlier stages, and then heating is required
to increase the temperature to match the observations. To match
the O VI bright knots, a higher heating rate is required such that
the heating energy is greater than the kinetic energy. The
temperatures for the O VI, Ly$\alpha$ and C III emission indicate
that different heating rates are required for knots bright in those lines. 
The required heating rates are much larger than
wave heating rates for the quiet Sun, and they
appear to be larger than predicted by thermal
conduction.  Magnetic heating should occur as
the expanding magnetic flux rope dissipates
energy to reach the simple configuration seen
in Interplanetary CMEs.  The flux rope model
of Kumar and Rust (1996) predicts that somewhat
more energy goes into heat than into kinetic
energy.  Our results for the knots bright in O VI
are compatible with that prediction, while the
heating rates for the knots bright in Ly$\alpha$
are smaller.

In \S2 we describe UVCS, LASCO, Extreme Ultraviolet
Imaging Telescope (EIT), and Transition Region and
Coronal Explorer (TRACE) observations. In \S3 we explain the
expansion model for reconstructing the three dimensional structure
from a time sequence of long slit spectra and show the three
dimensional structure of O VI, Ly$\alpha$, and C III emissions
observed by UVCS. In \S4 we describe the observational constraints
and the heating model. In \S5, the results are presented in terms of the
temperature evolution, heating energy, and energy balances. In \S6 we discuss our results and compare the heating rates with those expected for thermal conduction. In \S7 we summarize our studies.

\section{Observations}
UVCS \citep{kohl95} obtains spectra of the solar corona inside an
instantaneous field of view (FOV) given by the 40$'$ long
spectrometer entrance slits and the slit width (21$''$ in this
observation), which can be placed between 1.5 and 10 $R_\sun$. The
UVCS slit was pointed above NOAA 9973 (N16 E09 at 14:20 UT on Dec.
13 2001) at a position angle P.A.=349$^\circ$ at 2.4 $R_\sun$. The P.A. and height are the values at the point along the slit that is closest to Sun center, which is 5/9 of the way from the east edge of the slit. The series
of 300s exposures separated by about 30s readout time began many
hours before the event and lasted until 19:16 UT at the same
position. The data were obtained with a spatial binning
of 6 pixels (42$''$) and a spectral binning of 1 pixel, 0.0993
\AA\/ and 0.0915 \AA\/ for primary and redundant wavelengths,
respectively.

Figure 1 shows the composite image of EIT 195, UVCS, and LASCO C2
observations. A bright flare was observed in the EIT 195 \AA\/ band beginning at 14:24 UT, and the peak
intensity was seen at 14:36 UT in the 12 minute cadence images. In
this event, TRACE observed this active region in the UV continuum (1600 \AA\/ and 1700
\AA) and C IV 1550 \AA\/. A filament eruption was observed in the UV
continuum, 1600 \AA\/ around 14:20 UT (Figure 2). The leading edge
of the CME first appeared in the LASCO C2 FOV at 14:54 UT, and in
the C3 FOV at 16:18 UT. The {\it Geostationary Operational
Environmental Satellite} (GOES) X6.2 class flare began at 14:20 UT, peaked at
14:30 UT, and lasted until 14:35 UT. Figure 3 shows the observation time of each
instrument. EIT observed this active region
with 12 minute cadence. LASCO C2 and C3 obtain images with 24 minute cadence, however the C2 and C3 data were not available from 14:54 UT - 16:30 UT and 14:42 UT - 16:18 UT during the CME expansion, respectively. At 14:54 UT, C2 observed
the leading edge of this event and, then at the next observation, the CME has passed beyond of C2 FOV.

From 14:56 UT, UVCS observed Doppler shifted material in the O VI
1032 \AA, Ly$\alpha$ 1216 \AA$~$ and C III 977 \AA$~$ lines. Also the faint spectral lines, O V] 1218.35 \AA, and [O V] 1213.9 \AA, N III 989.90 \AA, N III 991.58 \AA, C II 1036.34 \AA, and C II 1037.02 \AA\ were observed in several exposures. It is especially important that the observed O V] ($2s^2 ~^1S_0 - 2s2p~^3P_1$) intercombination line (violating the selection rule $\triangle$ S=0, notation: right square bracket, O V]) at 1218.35 \AA\ and the [O V] ($2s^2~^1S_0 - 2s2p~^3P_2$) forbidden line (notation: square brackets, [O V]) at 1213.9 \AA\ can be used to determine the plasma density (see \S4.2.1).  Because the Einstein A value of the forbidden line is 0.022 $\rm s^{-1}$, the line is
quenched at densities above about $10^6~\rm cm^{-3}$ as shown in Figure 9 of Akmal et al. (2001). To investigate the heating rates, we analyzed the spectral lines for the fourteen blobs observed in four exposures (refer to Table 1). Figure 4 shows the UVCS observations of O VI (each left panel) and Ly$\alpha$, O V], [O V], N III (each middle panel), and C III (each right panel)
from 14:56 UT to 15:12 UT. The Ly$\alpha$ and O V lines are observed
in the redundant wavelength with the N III line in the primary
wavelength. The positions along the slit of the fourteen blobs are marked at the right axis of the middle panel with white bars. The intensities and line-of-sight velocites (V$_{LOS}$) of each blobs are presented in Table 2 (O VI and O V) and Table 3 (Ly$\alpha$, C III, C II, N III). For instance, in the case of blob A at the first detection of the CME at 14:56 UT, we selected the Doppler shifted O VI emission with -0.98 \AA\ and -1.68 \AA\ (evaluated by two Gaussian fit) indicating line-of-sight velocities of 286 km/s and 459 km/s, respectively at P.A. = 361 $^\circ$ - 366$^\circ$. In the exposure at 15:01 UT, we selected three blobs (B.1, B.2, and B.3) and those are also marked on the left panel of Figure 4. The C III emission of blobs A and B was detected at the edge of the observed wavelength range in this data panel.  In the case of region C observed at 15:07 UT, we selected six blobs, one detected only in O VI and C III (C.1), one detected in the bright emission of Ly$\alpha$ but relatively faint emission of O VI (C.2), and others detected in the emission of the forbidden [O V] $\lambda$1213.9 line (C.3 - C.6). In the 15:12 UT observations, we selected four blobs (D.1 - D.4) detected also in the emission of [O V]. The line intensities for most of blobs were evaluated by fitting two Gaussians to the line profile. 

\section{Three-dimensional reconstruction}
A three-dimensional reconstruction of the 13 December 2001 CME has
been made based on O VI 1032 \AA, Ly$\alpha$ 1216 \AA, and C III 977
\AA\ observed from 14:56 UT to 19:16 UT. Briefly, for each pixel along the UVCS slit, we first determine the velocity components in the plane of the sky for each exposure by dividing the difference between the eruption time and the time of the exposure into the difference in position between the pixel and the location of the eruption.  The observed Doppler shift provides the line of sight velocity component for each pixel in each exposure.  When different structures are superposed along the line of sight, they may show up as separate velocity components and can thus be separated into, for instance, structures on the front and back sides of the expanding CME \citep{lee06}, which were separated with multiple Gaussian fits.  We then use the 2 plane-of-the-sky velocity components and the line-of-sight velocity to project the structure forward in time, adding in the material observed in each successive exposure.  For this event, due to the 300 s exposure time,  blobs with different properties were sometimes superposed in the same pixel during  the same exposure.  The velocities in the plane of the sky would cause those blobs to be superposed at later times in the 2D projection onto the plane of the sky, but the different line-of-sight components of the blobs make it possible to separate them in the 3D structure.  Our method essentially assumes that the expansion is self-similar.  This assumption is checked by comparing the projection of the 3D structure onto the plane of the sky to LASCO images at later times. 
We use the same assumptions as the recent 3D reconstruction of
UV lines for the 2002, April 21 CME observed by UVCS by Lee et al. (2006) who give a detailed description of the method. This model has two key
assumptions. One is homologous expansion in all directions. The other
is that the line-of-sight velocities are only caused by the
material's expansion, rather than thermal or turbulent motions.

\subsection{Pre-processing of the data}
We subtracted as background the average of 35 pre-CME exposures
observed from 08:02 UT to 14:16 UT. The scattered Ly$\alpha$ continuum was
eliminated 
by a linear equation for the background emission in the observed wavelength ranges. The Gaussian wings on the instrumental line profile were corrected by an iterative procedure \citep{kohl97}. Since the
detector has some distortion, we evaluated the reference line center of
the O VI and Ly$\alpha$ by Gaussian fitting after binning by three spatial bins along the slit using the pre-CME observations
above. The line center of the C III was evaluated using the O VI line center.

\subsection{Expansion Model}
The onset position of the CME eruption is obtained from the position of
NOAA 9973, which was located at 0.084 $R_\sun$ to the west
(hereafter the x-coordinate) and 0.218 $R_\sun$ to the north (hereafter
the z-coordinate) according to the EIT image at 14:24 UT. The
position in the line-of-sight direction was thus -0.972 $R_\sun$
toward the Earth (hereafter the y-coordinate) assuming the Sun is a
sphere ($R_\sun$=1). First, We assume that each element of plasma moved in a straight line from the eruption site to the UVCS pixel where it was observed, as described above. Second, we assume that
all material erupts at the same time and location, and it has
constant speed before reaching the UVCS slit at each spatial bin.
This implies that the plasma reaching the slit during later
exposures travels more slowly in the plane of the sky than the
plasma that reaches the slit earlier. The ejection is spread out in both location and time, but only by about 100 arcseconds and 10 minutes in time as estimated from the TRACE movie.  Given the limitations of spatial resolution and cadence of the UVCS and LASCO data, the assumption of a single point in space and time is adequate.

\subsection{Images from the three-dimensional reconstruction}
A movie has been made using a 40 $R_\sun$ cube with 0.1 $R_\sun$
pixels (see electronic version). The reconstructed movie shows the distributions of the O
VI, Ly$\alpha$, and C III emissions (Figure 5). The red, green, and blue colors correspond to the O VI,
Ly$\alpha$, and C III emission, respectively. The yellow (O VI and
Ly$\alpha$), sky (Ly$\alpha$ and C III), violet (O VI and C III),
and white colors (O VI, Ly$\alpha$, and C III) show the overapping features of those lines. At the
beginning of the observation, 14:56 UT and 15:01 UT, the blue shifted components of C III emission were detected at the end of the wavelength range of this data panel (marked $^a$ in Table 3). This could underestimate the line-of-sight velocity to the Earth in the
reconstructed movie.

We present four views from the solar east-Earth (upper left), the solar east-backside of the Sun (upper right), the
backside of the Sun (lower left), and the solar west (lower right). The reconstructed structure is extended towards the Earth, as indicated by the strong blue shifts. The views from the
solar west (lower right) make it easy to understand the distribution of
the three emission lines along the line of sight. In the CME front, the O VI and Ly$\alpha$ emissions are positioned close together (red, green, and a mixed yellow colors). Following
those, the Ly$\alpha$ and C III emissions are closer to the middle
of the ejected CME plasma (green, blue, and a mixed sky colors). The
C III emission is seen in the innermost part of the reconstructed
movie, in agreement with earlier indications of a core of cooler material
\citep{ak01}. The view from the backside of the Sun to the Earth
(lower left panel in Figure 5) also shows the C III emission in the innermost part of the ejected CME material. The bright features of O VI, Ly$\alpha$, and CIII observed by UVCS show the arch (shell-like) structure in the view from the solar backside (please see the movie and the image at the bottom left in the Fig. 5). Inside of the shell-like structure, it is much fainter and seems to be void. Krall (2007) compared the CMEs observed by SMM with a parameterized 3D flux rope model \citep{kr06} and found that the CMEs observed by SMM in his study are consistent with the hollow flux-rope geometry.  Our 3D reconstruction looks like the hollow flux rope structure, as best seen in the view from the back of the Sun. 

\section{Energy Budget}

To investigate the energy balance of the 13 December 2001 CME plasma, we used the fourteen blobs from the 4 exposures observed by UVCS from 14:56:19 UT to 15:12:48 UT (Table 1, see also Figure 4). The bright features of the fourteen blobs occupy small spatial regions along the slit in one or more of the spectral lines. We analyze all the observed lines along the bright features which are marked with the white bar at the edge of the middle panel in Figure 4.

The procedure is much like that of Akmal et al. (2001), but for this event we have different constraints because different lines were observed, and because we use the three-dimensional reconstruction to constrain the depth along the line of sight. First, we evaluate the expansion of the CME plasma based on the TRACE, LASCO, and UVCS observations (\S4.1). Second, we compute grids of models for a broad range of initial temperatures, densities, and forms of the heating function to calculate the ionization fraction along the CME expansion. The ionization fraction is evaluated by a time-dependent ionization code using density and temperature as input which are calculated from the expansion law (\S4.2.2) and the energy equation (Eq. (5) in \S4.3). The ionization fractions are also used to compute the radiative cooling rate. At the final step of each model, the density, temperature, and- ionization fraction are used to evaluate the line intensities (Eq. (2)). Finally, the line intensities evaluated from each heating model are compared with the O VI, O V], Ly$\alpha$, C III, C II, and N III line emission observed by UVCS.

The heating laws for fast solar wind (Allen et al. 1998), in which proton and electron rates differ, as well as a density-proportional heating rate are used to model the heating rates.  The flux rope model for interplanetary magnetic clouds (Kumar and Rust 1996) is also used to evaluate the energy balances of the ejected plasma. The heating rates in the models contain all forms of thermal energy input including wave dissipation, magnetic heating and the divergence of thermal conduction, which we will consider in Section 6. Here we model the ejected plasma including radiative and adiabatic cooling to investigate the temperature evolution of the CME material with a time dependent ionization state model. We describe the model in more detail in the following subsections.

\subsection{Expansion Model}
We use an acceleration profile to evaluate the height of the ejected
plasma rather than the constant velocity that was used to make the three-dimensional reconstruction.
Chen et al. (2006) showed that the distance of the two footpoints in the active
region is related to the maximum expansion
velocity of the CME flux rope. We determined the peak time of the acceleration, 14:25:00 UT, based on the distance of the two footpoints from TRACE 1600 \AA\ observation. Then the start time of the expansion, 14:10:12 UT, was evaluated to match the heights of the CME plasma observed by TRACE, UVCS, and LASCO observations. Different slopes of the acceleration were used for each exposure. The applied accelerations are 900 $m/s^2$, 765 $m/s^2$ for the A and B blobs, respectively and 630 $m/s^2$ for the C and D blobs. The TRACE 1600 \AA\ observations also show that the brightening in the active region starts around 14:10 UT and increases slowly after that. The evaluated heights of the ejected CME plasma are presented in Figure 6.

\subsection{Observational Constraints}

Electron density, kinetic temperature, and line intensity are used to constrain the acceptable models for a broad range of initial temperature, density, and three heating functions for each of the fourteen blobs. The electron density evaluated at the UVCS slit height is used to determine the electron density at each time in the expansion model and to evaluate the emission measure (EM). The kinetic temperature of Ly$\alpha$ is used as the upper limit to the temperature at the UVCS slit height. The observed line intensities are compared with the intensities obtained from the models.

\subsubsection{Electron density}
The ratio of the [O V]/O V] lines is a reliable density diagnostic. This ratio has been used
to determine the electron density of the ejected CME plasma, and it is most useful for densities of 10$^6$-10$^7$ cm$^{-3}$ \citep{ak01}. In this event, UVCS observed the [O V] forbidden line
at 1213.85 \AA$~$ and the O V] intercombination line at 1218.39
\AA$~$ in several blobs, though the 1213.85 \AA$~$ line could not be cleanly separated from the N III 991.85 \AA$~$ line in most cases.  For the seven blobs observed at 15:07 UT and 15:12 UT, the electron number
densities were evaluated from [O V]/O V]. We present the line profiles of the three blobs C.5, C.6, and D.3 in Figure 7. Both primary and redundant wavelengths are presented together. The [O V] line may be contaminated by N III 991.58 \AA, and while we attempted to remove the N III by assuming that its profile is the same as that of C III 977 \AA, we consider the [O V] intensities to be upper limits and the density estimates to be lower limits. We use the electron number densities from the ratio of the [O V]/O V] for the three blobs, C.5, C.6, and D.3, for the anlysis of the heating rates and those are marked with $^\dag$$^\flat$ in Table 2. The electron density evaluated from the O V ratio and version 5.2 of CHIANTI \citep{de97, la06} is presented with uncertainties in Figure 8. For other
blobs, the electron number densities are evaluated from the column
densities observed by LASCO, divided by the line-of-sight depth based on the O VI line emission. The three dimensional reconstruction of the O VI lines shows the reconstructed width is similar to the CME width from the LASCO observation \citep{lee06}. We take the depth of the O VI line to be the FWHM multiplied by the travel time from the source region to the UVCS slit. The evaluated electron number density and column density were presented in Table 2.

The LASCO C3 observation at 16:18 UT was used to evaluate the column density
along the line-of-sight. Unfortunately LASCO didn't observe this event when the [O V] and O V] lines were observed from 15:07 UT to 15:12:48 UT (refer to Figure 3). The three-dimensional reconstruction makes it possible to estimate the height and the angle with the plane of the sky. The heights of the UVCS blobs predicted by the constant velocity reconstruction
     are higher than their positions in the LASCO C3 image at 16:18.  We use the positions from the reconstruction using the observed acceleration to connect the UVCS blobs to the LASCO image. In Figure 9, the reconstructed positions of the observed blobs are presented with the LASCO C3 observation, and each box represents the A, B, C, and D blobs in the Table 1. The reconstructed positions of blobs, A$-$D are also indicated by '3D position' in Figure 6. Vourlidas et al. (2000) showed how the densities derived from LASCO increase with
the angle with the plane of the sky due to the angular dependence of Thompson scattering.
Plane-of-the-sky angles of 40$^\circ$ $\sim$ 50$^\circ$
were found for the observed blobs from the line-of-sight velocities. We used the angle with the plane of the sky of 45 $^\circ$ to evaluate the column densities from the LASCO observation. The column densities were calculated by averaging the area of each box in Fig 9. After the column density was obtained, we scaled the density inversely with height squared to estimate the column density at the height of the UVCS observations.

\subsubsection{Temperature and line intensity}

The line widths obtained by the Gaussian fit for the fourteen blobs contain the kinetic temperature and non-thermal (bulk-velocity) components. The upper limits on the kinetic temperatures of O VI and Ly$\alpha$ are presented in Table 2 and Table 3, respectively. Note that in the 3-D reconstruction, the line width is assumed to be caused by the material's expansion along the line of sight. Here we regard the kinetic temperature to be the upper limit of the electron temperature at the UVCS slit height as one of the criteria for selecting the acceptable models. The Ly$\alpha$ width constrains only the ion temperature in the Ly$\alpha$ emitting region since the O VI and Ly$\alpha$ generally do come from different regions. The temperature $T_k$ is defined as:

\begin{equation}
T_k = \frac 1 2 \frac m {k} {v_{1/e}}^2
\end{equation}

\noindent where $m$ is the mass of the ion and v$_{1/e}$ is the velocity derived from Doppler half width, $\triangle\lambda_{1/e}$. We subtract about 0.3 \AA$~$ (3 pixels) in quadrature to account for the effect of the slit width of 0.298 \AA\/ (0.275\AA\/ for the redundant channel).

Radiative and adiabatic cooling along with the heating rates are used to investigate
the temperature evolution of the CME material in the time
dependent ionization state model. We use the expansion law, $n_{uvcs} /(n_0 - n_{uvcs}) \propto (t/t_{uvcs})^\alpha $, which determines the adiabatic cooling rate ($L_{adiabatic}$ in Eq. (5)) from $T/n^{\gamma-1}$ (Akmal et al. 2001, refer \S4 in their paper). Initial densities ($n_0$) are examined from $3 \times 10^8 cm^{-3}$ to $2 \times 10^{11} cm^{-3}$ with initial temperatures from $1 \times 10^4$ to $6.3 \times 10^6$. $t_{uvcs}$ is the duration time of the CME expansion from 14:10:12 UT (refer \S4.1) to the UVCS observations. Using the density ($n_{uvcs}$) determined at the UVCS slit, 2.4 $R_\sun$, the expansion is assumed to be a power law. We adopt the power law of $\alpha$=3, which means an expansion in both length and radius and a slower expansion at the earlier stages than the expansion with $\alpha$=1 or 2. Since we don't know exactly how the CME has expanded from the origin to the solar corona, the power law assumption which has been adopted in the earlier work (Akmal et al. 2001) is the simplest way to parameterize the density falloff with  height. Also the choice of the $\alpha$=3 matches the observed slower expansion in the earlier stages. During the expansion of the CME plasma, the ionization fraction is evaluated from the time-dependent ionization state model. The ionization and recombination rates were taken from an updated version of the code of Raymond (1979). This atomic physics package computes the collisional ionization, radiative recombination and dielectronic recombination rates for the astrophysically abundant elements, and the evolution of the ionization states of the elements was computed with the from the usual set of equations for the rate of change of each ionic fraction. The time-dependent ionization balance was used with solar photospheric abundances \citep{gr98} to compute the radiative cooling rate ($L_r$ in Eq. (5)). Photospheric abundances were chosen because the knots we analyze are probably ejected prominence material, rather than coronal plasma. The computed ionization fraction is used to determine the line intensity. In coronal conditions with the isothermal assumption, the line intensity is expressed as follows (e.g. Ko et al. 2006).

\begin{equation}
I_{line}=\frac 1 {4\pi} \frac {n_{el}} {n_{H}} \int \epsilon(T_e)dEM(T_e) ~~~~~ photons ~ cm^{-2} ~ s^{-1} ~ sr^{-1}
\end{equation}

\noindent Where $n_{el}/n_H$ is the abundance of the element relative to hydrogen. In our model, the line intensity is calculated for a single temperature at any time. So we use the emission measure, $EM$ ($=0.8 n_e N_e$) instead of $dEM$ without the integral. 0.8 is the hydrogen abundance relative to the electron density, $n_e$ is the electron number density determined from the O V lines or from LASCO observation divided by line-of-sight depth (see \S4.2.1), and $N_e$ is the column density determined from the LASCO image (see \S4.2.2). $\epsilon(T_e)$ is the emissivity, which is defined as

\begin{equation}
\epsilon(T_e) = \frac {n_{ion}} {n_{el}} (T_e) q_{line}(T_e),
\end{equation}

$n_{ion}/n_{el}$ is the ionization fraction. q$_{line}$ is the collisional electron excitation rate given by

\begin{equation}
q_{line}(T_e) = \frac {8.63\times 10^{-6}} {\sqrt{T_e}} \frac {\Omega_{ij}} {w_i} e^{-E/kT_e},
\end{equation}

where $\Omega_{ij}$ is the collision strength taken from version 5.2 of CHIANTI and $w_i$ is the statistical weight.

The observed line intensities are compared with the line intensities predicted from the three heating models (see \S 4.3). To match the O VI blobs, we use the ratio of the O VI/ O V] which allows a factor of two margin because the density obtained from O V may not pertain to the O VI emitting gas. To match the Ly$\alpha$, C III, C II, and N IIII blobs, we require that the predicted emission is at least as large as the observed values, but it is smaller than three times the observed intensity. We allow this margin because of the difficulty in evaluating the emission measure, given that we do not have a density from the O V lines for this cooler gas, and the combination of line-of-sight depth with LASCO column density carries a greater uncertainty. When the C III was observed near the edge of the panel, we exclude this constraint. In cases where the O VI/O V] ratio is not well matched (C.4 and D.4 for the heating model of the Kumar and Rust, marked by $^c$ in Table 4), we use the same constraints for the Ly$\alpha$, C III, C II, and N III blobs. We present the detailed constraints and the evaluated energy in Table 4.

\subsection{Heating model}
Earlier UVCS observations of CMEs found evidence that the ejected plasma is heated
long after the eruption \citep{ak01, cia01}. We investigate the heating rate for the following three heating models (refer to \S4.3.1 and \S4.3.2). The heating rate, H, enters the energy equation,

\begin{equation}
\frac5 2 n k \frac {dT} {dt} = -n_e n_p (L_r + L_{adiabatic} - H )
\end{equation}

\noindent where $n$ is the density of both protons and electrons. k is the Boltzmann constant and $n_e$ and $n_p$ are an electron and proton density, respectivly. $L_r$ and $L_{adiabatic}$ are radiative cooling rate and adiabatic cooling rate, respectively. H includes all contributions to the heating, such as wave dissipation, magnetic reconnection, turbulent heating or the divergence of conductive flux, or shock waves generated by the reconnection outflow \citep{sh05}.  We combine turbulent heating with wave dissipation, so

\begin{equation}
H = H_{wave} + H_{mag} + H_{cond} + H_{shock}
\end{equation}

We have used the physical value of the specific heat ratio $\gamma$=5/3. Smaller values of $\gamma$ are often used in coronal models
with the heating rate set equal to zero in order to obtain plausible temperature profiles without knowledge of the heating rate.  In this case we aim to derive the heating rate, so the value of $\gamma$ for a monatomic gas is appropriate.
 
We use a wide range of heating rates with three forms of the heating function. Two of them are taken from parameterized heating rates for the solar corona, and the third is from the CME model of Kumar and Rust.

\subsubsection{Parameterized heating}
The heating law for fast solar wind (Allen, Habbal and Hu, 1998), in
which proton and electron heating rates differ (hereafter denoted as Q $\propto$ Q$_{AHH}$), and the density-proportional heating (hereafter Q $\propto$ n) are used to evaluate the heating rates.

The density proportional model was chosen, as in Akmal et al. (2001), to provide a heating rate that declines
more gradually than the exponential rate of Allen et al. (1998). It implies no specific physical mechanism, but for instance
saturated thermal conduction with a constant temperature gradient might give such heating.  Shiota et al. (2005)
have described how the impact of reconnection outflows on the trailing edge of a CME core can generate slow
mode shock waves that travel around and into the core. Lacking knowledge of the evolution of the reconnection
outflow or the location of the shock dissipation, we cannot predict the functional form of this heating rate,
but it is plausible that it declines on a time scale comparable to the travel time to the height of the UVCS slit.

For the model, Q $\propto$ Q$_{AHH}$, the proton and electron heating rates (Q$_p$ and Q$_e$) are defined as,

\begin{equation}
Q_e = Q_{e0} e^{-(r-R_S)/\sigma_e} \qquad,
Q_p = Q_{p0} e^{-(r-R_S)/\sigma_p} + q00\left( \frac {R_S} {r} \right)^4 (1-e^{-(r-R_S)/5R_S}),
\end{equation}
in their paper. We consider a constant heating rate with an exponential falloff with scale height 0.7 $R_\sun$ ($\sigma_{e,p}$). Q$_{e0}$, Q$_{p0}$, and $q00$ are the strength of heating terms taken from the SW3 model in their paper. $r$ is the distance from the initiation site and R$_S$ is solar radius. The temperature changes due to the proton and electron interaction are considered for the heating model, Q $\propto$ Q$_{AHH}$. We model the heating functions for a range of constants of proportionality covering three orders of magnitude.

\subsubsection{Magnetic Heating}
The magnetic clouds in Interplanetary Coronal Mass Ejections generally have simple magnetic structures that can be
fit with a Lundquist solution (e.g. Lynch et al. 2005), a minimum energy state for a given magnetic helicity.  To
reach this state, the complex, stressed magnetic configuration of the erupting structure must relax by dissipating magnetic
energy. The MHD model of Lynch et al. (2004) shows that this occurs by the time the CME reaches about 15 $R_\sun$.
MHD models do not generally show predicted heating rates, and in any case the heating is unlikely to be uniform throughout the
CME structure. However, the model of Lynch et al. (2008) indicates that about 15\% of the total change in magnetic free energy
goes into kinetic energy.

In order to parameterize magnetic heating we use the flux rope
model for interplanetary magnetic clouds of Kumar and Rust
(1996; hereafter $Q \propto Q_{KR}$). The model predicts that the magnetic energy goes
into overcoming solar gravity, kinetic energy for the CME expansion
and heating energy. An important feature of the model is its use of the principle of conservation of magnetic helicity. It suggests that the magnetic energy stored in an expanding plasma should decrease with expansion, ($H_m \propto$ length scale $\times$ $U_m$), where $H_m$ is the magnetic helicity and $U_m$ is the magnetic energy. Kumar and Rust (1996) assume that the flux rope evolves through a self-similar series of axisymmetric states each having the lowest energy consistent with conservation of magnetic helicity.  The assumption of self-similar axisymmetric expansion may be adequate when the CME is much larger than the source region \citep{ur05}, but it is not likely to be accurate at small heights \citep{ak01}. Kumar and Rust (1996) also assume that magnetic energy is dissipated in the expansion process, though they do not identify the dissipation mechanism in detail. 

We use the length scale, $l = 2\pi a$, where $a$ is the radius of the flux rope (refer to Figure 2 in their paper). We use the distance of the CME plasma from the initiation site as the radius 'a'. Kumar and Rust found that the change of magnetic energy is $dU_m = U_m(l_0)l_0 dl/l^2$ and that as the CME expands the magnetic energy lost is partitioned between kinetic + gravitational and thermal energies according to a parameter $s = 1 - sin(\theta_0)/\pi$  that may vary from one event to another. $\theta_0$ is an effective toroidal angle (refer Fig 5 in their paper). As stated in their paper, the kinetic energy of major radial expansion about the center of mass is same as the kinetic energy of center of mass (the kinetic energy of the minor radial expansion is negligible for a large aspect ratio toroidal cloud) and the force acting on the center of mass is maximum on the leading edge of the flux rope. With twice the center of mass kinetic energy derived by integrating of the motion of the center of mass from the initial distance to the leading edge of the flux rope ($\theta_0$ $\leqq$ $\pi$/2), they presented energy conservation as $dQ + dU_m + dU_G + dU_{KE}=0$. Using the energy conservation, we derive the heating energy during the expansion of the CME plasma. If $dU_G$ is the change in gravitational energy and $dU_{KE}$ is the change in kinetic energy, then

\begin{equation}
dQ = - hdU_m
\end{equation}

\noindent where $h=(1+dU_G/dU_m+dU_{KE}/dU_m)$ is the fraction of the lost magnetic energy appearing as heat. We find the kinetic and gravitational energy from the velocities and distances explained in \S 4.1. The final kinetic energy and the initial gravitational energy are fixed by the observed velocity and an initial height of CME plasma, respectively. We find the heating rates to match the observations for $\theta_0$ in the range of 10$-$90 $^\circ$. The blobs we observe appear to be part of the ejected prominence, and as such they are part of flux rope. The energy budgets we obtain pertain specifically to these blobs, but lacking any reason to assume otherwise, we believe that the heating is uniform throughout the flux rope.

\section{Energy balances}

\subsection{Temperature evolution}

Figure 10 shows the temperature evolution of both protons and electrons when one of the acceptable heating rates is applied for matching the O VI and O V] line intensities of blob C.5. The applied heating rate is 63 times that for the fast solar wind model (Q$\propto$Q$_{AHH}$). The upper two panels show the proton and electron temperature evolutions for all applied initial densities (3.16$\times$10$^8$$-$1.99$\times$10$^{11}$ cm$^{-3}$) for the particular initial temperature, 1.58$\times$10$^6$ K. The higher initial densities (red and yellow ranges in the rainbow colors) drop more rapidly than the lower initial densities, but all initial densities reach the same temperature at the time of the UVCS observation. The lower two panels show the same feature as that above for all applied initial temperatures (1.$\times$10$^4$ $-$ 6.3$\times$10$^6$ K) at that particular initial density, 5$\times$10$^8$ cm$^{-3}$. The proton temperature increases more than the electron temperature because of the different heating rates for the protons and electrons (see \S4.3.1). The temperatures drop rapidly at the lower heights and in the earlier stages of the expansion, and then they remain fairly constant depending on the heating rates. This is a similar result to that of Akmal et al. (2001) (see \S5.2 and Figure 12 in this paper, also see Figure 12 in their paper). Our models show that the temperatures at coronal heights are determined by the heating rates rather than the initial temperatures and densities. In Figure 11, we present the line intensities (upper panels) and the line ratios (lower panels) for all initial densities with the temperature predicted from a range of heating rates which are 0.63
- 630 times Q$_e$ and Q$_p$ of the Q$\propto$Q$_{AHH}$ (initial temperature = 1.58$\times$10$^6$ K). The temperatures that agree with the O VI, Ly$\alpha$, C III, C II, and N III emissions indicate that different heating rates are required for those ions. For this model, the heating rates for the O VI blobs and for the blobs of cooler ions are 63 times and 2.5 times for the fast solar wind model, respectively. Hereafter H1 represents the heating rates that match the O VI emission and H2 represents the heating rates for matching the Ly$\alpha$, C III, C II, and N III emission.

\subsection{Heating, cooling, and energy budget}

Figure 12 shows the evolution of the temperature, density, and ionization state for one of the acceptable models explained above. The temperature and the ionization fractions change rapidly at the lower heights. For the H1 models, the O VI and O V ionic fractions maintain higher values after the rapid increase at lower heights, but the fractions of other ions, Ly$\alpha$, C III, C II, and N III emissions, decrease again (the middle panel in Figure 12). For the H2 model, the fractions of the O VI and O V decrease and the others maintain their fractions (the right panel in Figure 12). The rapid change in the ionic fration at low heights reflects the change in the eletron temperature when they are in ionization equilibrium. On the contrary, at heights higher than 2 R$_\sun$, the rapid expansion tends to 'flatten' the ionic fraction indicating that the ions are frozen-in. The outflowing ions flow too fast and the density is too low for them to exchange electrons between neighboring ionization states from collisions. Figure 13 shows the heating and cooling rates for the H1 model in the upper panels. The adiabatic cooling rate increases rapidly at the earlier stages (the upper right in Figure 13). This drastic cooling rate requires that continuous heating increases the temperature to match the observations. The thermal and radiative loss energies are smaller than the heating and adiabatic cooling energies. The heating energies for the acceptable models for each of the three heating functions are shown in Figure 14. The heating rates for the O VI  (H1) and for the Ly$\alpha$, C III, C II, and N III (H2) observations are presented as diamonds and crosses, respectively. H1 is over ten times higher than H2 and the Q$\propto$Q$_{KR}$ model does not match the observations for the Ly$\alpha$, C III, C II, and N III blob. The integrated heating energy balances cooling energy, and the total heating is greater than the kinetic energy in both the Q$\propto$Q$_{AHH}$ and Q$\propto$Q$_{KR}$ models, as shown in the lower right in Figure 13 and Figure 15. The Kumar and Rust model predicts that the magnetic energy decreases conserving magnetic helicity.

\section{Discussion}
In Table 4, we present the heating and thermal energy for all fourteen blobs. The kinetic energy of this CME is 1.6-3.3$\times$10$^{15}$ erg/g which is evaluated from the velocity of the CME plasma at the UVCS slit, 2.4 R$_\sun$. The gravitational and ionization energy are 5.6$\times$10$^{14}$ erg/g and 1.9$-$2.1$\times$10$^{13}$ erg/g, respectively. In cases where the model does not satisfy the constraints from the observations, these are marked as a - e (details in Table 4). The H2 model could not satisfy all the constraints together for the several blobs because the lines Ly$\alpha$, C III, C II, N III might come from different regions along the line of sight. For all the acceptable models, the heating energy for H1 is higher by nearly one order of magnitude than that for H2, and it exceeds the kinetic energy. A recent study of heating rates for Interplanetary coronal mass ejection (ICME) from Helios 1 and Helios 2 observations found that plasma turbulence heats the proton and alpha particles and determined the heating rates to be 1.9 $\times$10$^{8}$erg g$^{-1}$ s$^{-1}$ at 0.3 AU and 1.9 $\times$10$^{5}$erg g$^{-1}$ s$^{-1}$ at 20 AU \citep{liu06}. The heating rate, 1$\times$10$^{11}$ erg g$^{-1}$ s$^{-1}$ at 0.01 AU, estimated by extrapolating from their study (Figure 6, in their paper) is similar to our result of the heating rate, ~5$\times$10$^{-7}$ erg cm$^{-3}$ sec$^{-1}$ at 2.4 R$_\sun$ (taking the electron density 1.$\times$10$^6$ cm$^{-3}$ and the mass from assuming 10 $\%$ Helium, 1.974$\times$10$^{-24}$ g).

The heating energies from the Kumar and Rust model are also greater than the kinetic energy. For all acceptable models, 70-90$\%$ of the magnetic energy goes into the heating energy, in agreement with their theoretical calculation. The December 13 2001 event is probably a prominence eruption and it is presumably defined by magnetic field (refer to Figure 2). The fractions of the magnetic energy that go into heat for the earlier observed blobs are higher than these later ones. For H2, we couldn't find a heating rate that satisfies the intensities of those low-temperature lines. It is possible that the CME plasma inside the flux rope does not follow the heating law of the Kumar and Rust magnetic cloud model or that the assumed self-similar expansion is not a good approximation in the early stages of the eruption.

Figure 16 shows the time scale of radiative cooling, adiabatic cooling, and thermal conduction for one of the acceptable models for H1 of C.5 (Q $\propto$ n). We can use the temperature gradients from this model with the Spitzer thermal conductivity to evaluate the importance of the thermal conduction contribution to the heating.  Larger temperature gradients would give too low a temperature at the  height of the UVCS slit, so this gives an upper limit to the thermal conduction contribution.  Other parameterizations of the heating rate could differ slightly, but not at the level needed for thermal  conduction to balance adiabatic cooling.   
The thermal conductive flux and time scale can be estimated as

\begin{equation}
\noindent \nabla F_c = -\kappa_0 \left[ \left( \frac {T_{i}+T_{i-1}} {2} \right)^{5/2} \frac {(T_{i}-T_{i-1})} {(l_{i}-l_{i-1})}
	      -\left( \frac {T_{i+1}+T_{i}} {2} \right)^{5/2} \frac {(T_{i+1}-T_{i})} {(l_{i+1}-l_{i})} \right] / (l_{i+1}-l_{i-1})
\end{equation}

\noindent and

\begin{equation}
\tau_i = \frac{\frac5 2 n_{i-1} k T_{i-1}}  {\nabla F_c},
\end{equation}

\noindent respectively. Here {\it i} is time step and $\nabla F_c$ is the divergence of the conductive flux, $\kappa_0$ is 1.8$\times$10$^{-5}$/$\ln\Lambda$ erg cm$^{-1}$ K$^{-7/2}$ sec$^{-1}$, and $\ln\Lambda$ is Coulomb logarithm. In the earliest stage of the expansion, the time scale of thermal conduction is much smaller than other cooling time scales. In this stage, the thermal conduction would maintain the initial high temperature and might change the temperature structure during the first 200 seconds and flatten out the initial steep drop(or rise) in temperature if the temperature gradient lies along the magnetic field. In the plot of the thermal conduction time scale, the peak around 300 seconds is due to the change of the sign of the divergence of the conductive flux. Those plots indicate that the adiabatic cooling is much more important than other terms after earlier stages. We have assumed classical Spitzer conductivity in equation (9).  Saturated conductivity would provide a smaller heating rate, but it is possible that a non-Maxwellian velocity distribution gives a larger conductive heating rate.

Due to the expansion of CME, the plasma cools rapidly at earlier stages, and then the heating energy becomes the major contribution in the energy budget of the ejected CME plasma, even greater than the kinetic energy. Guhathakurta et al. (2006) studied the heating of the corona and they also found that the thermal and non-thermal heating terms contribute to the temperature and heat flux in the low corona (refer Figure 4 in their paper). A similar result for the heating energy of CME plasma based on the Advanced Composition Explorer (ACE) observations shows that the plasma requires further heating following filament eruption \citep{ra07}.

Our results confirm the studies of Akmal et al. (2001) for a different event and obtain a narrower range of the heating rates in both ions and electrons. Also our study shows that heating rates are consistent with Kumar and Rust's model. About 75 $\%$ of the magnetic energy goes into the heating energy for the O VI observation with Kumar and Rust's model, in agreement with their prediction.

\section{Summary}

UVCS observed Doppler shifted material in the O VI 1032 \AA, Ly$\alpha$ 1216 \AA, and C III 977 \AA\ lines on December 13 2001. Fainter spectral lines, O V] 1218.35 \AA, [O V] 1213.9 \AA, N III 991.58 \AA, C II 1037.02 \AA, were observed in several exposures. A three-dimensional reconstruction of the 2001 December 13 partial halo CME has been made based on the O VI, Ly$\alpha$, and C III lines. The reconstructed structure shows the distribution of those ions as viewed from all directions. The structure viewed from the backside of the Sun towards the Earth looks like a hollow flux rope. The C III emission is located in the innermost part of the ejected CME material. The reconstructed structure is also used to find the column density with LASCO observations. We investigate the heating rates by a procedure similar to that of Akmal et al. (2001). For the individual knots observed by UVCS, we generate a large grid of models having different initial conditions and forms of the heating function. We find that continuous heating is required to match the UVCS observations. The temperature evolution shows a rapid decrease at lower heights and earlier stages, and then heating is required to increase the temperature to match the observations. To match the O VI bright knots, a higher heating rate is required such that the heating energy is greater than the kinetic energy. The temperatures for the O VI and Ly$\alpha$, C III, C II, N III emission indicate that different heating rates are required for the bright knots in those lines. About 75 $\%$ of the magnetic energy goes into the heating energy for the O VI observation with Kumar and Rust's model, in agreement with their prediction, but those models do not match the cool line blobs.

\acknowledgments

We thank G.-S. Choe and Y.-J. Moon for the helpful discussion on this paper. This work was supported by the Korea Research Foundation (ABRL-R14-2002-043-01001-0) and the Grant NNG06GG78G to the Smithsonian Astrophysical Observatory. J.-Y. Lee acknowledges the support of the Korea Research Foundation (KRF-2005-070-C000059), the Astrophysical Research Center for the Structure and Evolution of the Cosmos (ARC-SEC) of Korea Science and Engineering Foundation through the Science Research Center (SRC) program, and the Grant NNM07AA02C to the Smithsonian Astrophysical Observatory.

\clearpage

\begin{deluxetable}{cccl}
\tabletypesize{\scriptsize}
\tablecaption{Blobs observed by UVCS}
\tablewidth{0pt}
\tablehead{
\colhead{Blob No.} & \colhead{UT} & \colhead{P.A.} & \colhead{Observed Lines}
}

\startdata
A   & 14:56:19 & 361\arcdeg-366\arcdeg & O VI, Ly$\alpha$, C III, C II $\lambda$1036.34, C II $\lambda$1037.02, N III $\lambda$991.58 \\
B.1 & 15:01:47 & 363\arcdeg-364\arcdeg & O VI, Ly$\alpha$, C III, C II $\lambda$1036.34, C II $\lambda$1037.02, N III $\lambda$991.58 \\
B.2 &          & 360\arcdeg-361\arcdeg & O VI, O V], Ly$\alpha$, C II $\lambda$1037.02, N III $\lambda$991.58 \\
B.3 &          & 350\arcdeg-352\arcdeg & O VI, O V], Ly$\alpha$, C III, C II $\lambda$1036.34, C II $\lambda$1037.02, N III $\lambda$991.58 \\
C.1 & 15:07:18 & 357\arcdeg-358\arcdeg & O VI, C III \\
C.2 &          & 350\arcdeg-352\arcdeg & O VI, O V], Ly$\alpha$, C III, N III $\lambda$ 989.90, N III $\lambda$991.58 \\
C.3 &          & 347\arcdeg             & O VI, O V], [O V], Ly$\alpha$, C III, C II $\lambda$1037.02, N III $\lambda$991.58 \\
C.4 &          & 346\arcdeg             & O VI, O V], [O V], Ly$\alpha$, C III, C II $\lambda$1037.02, N III $\lambda$991.58\\
C.5 &          & 345\arcdeg             & O VI, O V], [O V], Ly$\alpha$, C III, C II $\lambda$1037.02, N III $\lambda$991.58\\
C.6 &          & 344\arcdeg             & O VI, O V], [O V], Ly$\alpha$, C III, C II $\lambda$1037.02, N III $\lambda$991.58\\
D.1 & 15:12:48 & 347\arcdeg             & O VI, O V], Ly$\alpha$, C III, N III $\lambda$991.58\\
D.2 &          & 346\arcdeg             & O VI, O V], [O V], Ly$\alpha$, C III, C II $\lambda$1036.34, C II $\lambda$1037.02, N III $\lambda$991.58 \\
D.3 &          & 345\arcdeg             & O VI, O V], [O V], Ly$\alpha$, C III, C II $\lambda$1036.34, C II $\lambda$1037.02, N III $\lambda$991.58\\
D.4 &          & 344\arcdeg             & O VI, O V], [O V], Ly$\alpha$, C III, N III $\lambda$991.58
\enddata

\end{deluxetable}

\begin{deluxetable}{clcccccccccc}
\tabletypesize{\scriptsize}
\setlength{\tabcolsep}{0.02in}
\tablecaption{Line intensities of the O VI, O V, and electron density}
\tablewidth{0pt}
\tablehead{
\colhead{Blob No.} & \multicolumn{6}{c}{O VI} & \multicolumn{2}{c}{O V]} & \multicolumn{2}{c}{ne($\times$10$^6$)} & \colhead{}{Ne($\times$10$^{16}$)} \\

\colhead{} & \colhead{I$_{1032}$} & \colhead{V$_{LOS}$} & \colhead{V$_{1/e}$} & \colhead{T$_k$} & \colhead{Depth} & \colhead{I$_{1032}$/I$_{1037}$} & \colhead{I$_{1218}$ } & \colhead{I$_{1213}$/I$_{1218}$} & \colhead{O V} & \colhead{LASCO} &  \colhead{LASCO}
}

\startdata
A &  34.0$^*$ & -286 & 58 & 33 & 0.858 & 2.75$\pm$0.14 & \nodata & \nodata & \nodata & 2.04{$\flat$} & 13.6 \phn \\
&39.4$^*~\chi^2$=0.9 & -459 & 97 & 91 & & & & & & &  \\

B.1 & 261.6$^*$ & -286 & 81 & 63 & 1.217 & 2.43$\pm$0.15 & \nodata & \nodata & \nodata & 1.31{$^\flat$} & 12.4 \phn \\
&132.0$^*~\chi^2$=3.7 & -459 & 112 & 121 & & & & & & & \\

B.2 & 307.7$~\chi^2$=0.9 & -487 & 58 & 33 & 0.367 & 2.78$\pm$0.19 & 70.8$\pm$7.0 & \nodata & \nodata & 5.14{$^\flat$} & 14.6 \phn \\

B.3 & 129.1$~\chi^2$=2.8 & -260 & 90 & 78 & 0.566 & 2.05$\pm$0.15 & 37.5$\pm$5.1 & \nodata & \nodata & 2.38{$^\flat$} &  14.7 \phn \\

C.1 & 81.3$~\chi^2$=0.2 & -457 & 56 & 30 & 0.394 & 3.03$\pm$0.32 & \nodata & \nodata & \nodata & 3.21{$^\flat$} & 9.68 \phn \\

C.2 & 118.8$~\chi^2$=3.1 & -346 & 57 & 31 & 0.401 & 1.86$\pm$0.13 & 35.6$\pm$4.9 & \nodata & \nodata & 5.59{$^\flat$} & 17.1 \phn \\

C.3 & 574.7$~\chi^2$=4.8 & -263 & 71 & 49 & 0.501 & 2.53$\pm$0.09 &
62.5$\pm$6.5 & 0.57$\pm$0.11 & 1.28{$^{+0.49}_{-0.35}$}{$^\dag$} & 5.88{$^\flat$} & 22.4 \phn \\

C.4 & 621.4$~\chi^2$=14.8 & -234 & 87 & 73 & 0.613 & 2.03$\pm$0.07 &
39.5$\pm$5.2 & 0.57$\pm$0.21 & 1.25{$^{+1.08}_{-0.53}$}{$^\dag$} & 5.23{$^\flat$} & 24.3 \phn \\

C.5 & 71.1$^*$ & -66 & 38 & 14 & 0.652 & 1.98$\pm$0.07 &
49.5$\pm$5.8 & 0.47$\pm$0.12 & 1.64{$^{+0.81}_{-0.46}$}{$^\dag$}{$^\flat$} & 5.73 &  28.3 \phn \\
&439.2$^*~\chi^2$=3.6 & -239 & 54 & 28 &  &  &
&  &  &  \phn \\

C.6 & 100.8$^*$ & -95 & 71 & 49 & 0.760 & 2.05$\pm$0.10 &
95.2$\pm$8.1 & 0.10$\pm$0.04 & 11.2{$^{+10.9}_{-3.74}$}{$^\dag$}{$^\flat$} & 3.67 & 21.1 \phn \\
&184.0$^*~\chi^2$=2.0 & -239 & 36 & 13 &  &  &
&  &  &  \phn \\

D.1 & 176.3$~\chi^2$=7.5 & -350 & 68 & 45 & 0.537 & 2.34$\pm$0.14 &
54.7$\pm$6.1 & \nodata & \nodata & 3.70{$^\flat$} & 15.0 \phn \\

D.2 & 167.3$~\chi^2$=10.8 & -321 & 48 & 22 & 0.378 & 1.83$\pm$0.14 &
43.3$\pm$5.4 & 0.63$\pm$0.17 & 1.06{$^{+0.71}_{-0.39}$}{$^\dag$} & 6.39{$^\flat$} & 18.2 \phn \\

D.3 & 142.0$~\chi^2$=4.5 & -297 & 30 & 8.7 & 0.234 & 2.79$\pm$0.22 &
86.7$\pm$7.7 & 0.30$\pm$0.06 & 3.08{$^{+0.88}_{-0.72}$}{$^\dag$}{$^\flat$} & 1.23 & 19.9 \phn \\

D.4 &  82.3$~\chi^2=$3.8 & -297 & 45 & 20 & 0.350 & 3.43$\pm$0.42 &
52.5$\pm$6.0 & 0.37$\pm$0.10 & 2.31{$^{+1.11}_{-0.64}$}{$^\dag$} & 7.35{$^\flat$} & 19.4 \phn \\

\enddata
\tablecomments{\\
               I : Intensity (10$^8$ photons/cm$^2$ sec sr)\\
               V$_{LOS}$, V$_{1/e}$ : km / sec\\
               T$_k$ : 10$^5$K ~~ \\
               Depth : Line of sight depth in R$_\sun$ unit evaluated by (FWHM of the O VI) $\times$ (travel time from the source region to UVCS slit) \\
              {$^*$} : Two Gaussian fit\\
              {$^\dag$} : The lower limit of electron density evaluated by [O V]/O V]\\
              {$^\flat$} : The adapted electron number density for heating model
}

\end{deluxetable}

\begin{deluxetable}{clccclcccccc}
\tabletypesize{\scriptsize}
\setlength{\tabcolsep}{0.02in}
\tablecaption{Line intensities of the Ly$\alpha$, C III, C II, and N III}
\tablewidth{0pt}
\tablehead{
\colhead{Blob No.} & \multicolumn{4}{c}{Ly$\alpha$} & \multicolumn{3}{c}{C III} & \multicolumn{2}{c}{C II (Int.)} & \multicolumn{2}{c}{N III (Int.)}  \\
\colhead{}
& \colhead{I$_{Ly\alpha}$} & \colhead{V$_{LOS}$} & \colhead{V$_{1/e}$} & \colhead{T$_k$}
& \colhead{I$_{C III}$} & \colhead{V$_{LOS}$} & \colhead{V$_{1/e}$}
& \colhead{1036.34 \AA\ } & \colhead{1037.02 \AA\ }
& \colhead{989.80 \AA\ } & \colhead{991.58 \AA\ }
}

\startdata
A & 4351.8$~\chi^2$=222.4 & -380 & 90 & 4.9
& 142.6$^a\pm$5.5 & \nodata & \nodata
& 2.6$\pm$0.7 & 3.2$\pm$0.8 &
\nodata & 48.9$^c\pm$3.2 \phn \\

B.1 & 1776.9$~\chi^2$=72.7 & -425 & 64 & 2.5
& 334.3$^a\pm$8.5 & \nodata & \nodata
& 7.7$\pm$1.3 & 20.5$\pm$2.1 &
\nodata & 42.0$\pm$1.4 \phn \\

B.2
& 786.7$\pm$23.2 & \nodata & \nodata & \nodata
& \nodata$^a$ & \nodata & \nodata
& \nodata & 9.0$\pm$1.4
&\nodata & 48.4$\pm$3.2 \phn \\

B.3
& 5649.5$~\chi^2$=51.8 & -325 & 80 & 3.9
& 397.6$\pm$9.3 & \nodata & \nodata
& 5.8$\pm$1.1 & 20.5$\pm$2.1
&\nodata & 23.8$\pm$2.3 \phn \\

C.1
& \nodata & \nodata & \nodata & \nodata
& 14.5$\pm$1.8 & \nodata & \nodata
& \nodata & \nodata
& \nodata & \nodata \phn \\

C.2
& 2169.3$~\chi^2$=4.1 & -393 & 70 & 3.0
& 285.2$^a\pm$7.8 & \nodata & \nodata
& \nodata & \nodata
&6.3$\pm$1.2 & 20.5$\pm$2.1 \phn \\

C.3
& 1226.5$^*$ & -344 & 56 & 1.9
& 220.5$^*$ & -132 & 30
& \nodata & 12.6$\pm$1.6
& \nodata & 26.0$\pm$2.4 \phn \\
& 487.3$^*~\chi^2$=34.1 & -231 & 35 & 0.74
& 195.6$^*~\chi^2$=4.0 & -193 & $^b$
&&&& \phn \\

C.4
& 1254.3$^*$ & -321 & 67 & 2.7
& 40.4$^*$ & -41 & $^b$
& \nodata & 9.3$\pm$1.4
& \nodata & 18.1$\pm$2.0 \phn \\
& 891.9$^*~\chi^2$=22.0 & -208 & 49 & 1.5
& 239.1$^*~\chi^2$=7.4 & -193 & 17
&&&& \phn \\

C.5
& 337.5$^*$ & -317 & 41 & 1.0
& 72.6$^*$ & -46 & 18
& \nodata & 12.2$\pm$1.6
& \nodata & 29.3$\pm$2.5 \phn \\
& 1291.2$^*~\chi^2$=31.9 & -136 & 45 & 1.2
& 226.1$^*~\chi^2$=3.7 & -198 & $^b$
&&&& \phn \\

C.6
& 425.9$^*$ & -317 & 48 & 1.4
& 75.4$^*$ & -76 & 31
& \nodata & 5.0$\pm$1.0
& \nodata & 21.4$\pm$2.1 \phn \\
& 565.0$^*~\chi^2$=30.9 & -114 & 35 & 0.74
& 190.4$^*~\chi^2$=11.4 & -198 & $^b$
&&&& \phn \\
D.1
& 1442.6$~\chi^2$=9.6 & -366 & 39 & 0.92
& 285.8$\pm$7.9 & \nodata & \nodata
& \nodata & \nodata
& \nodata & 21.8$\pm$2.2 \phn \\

D.2
& 1441.0$~\chi^2$=11.6 & -344 & 37 & 0.83
& 178.7$\pm$6.2 & \nodata & \nodata
& 3.8$\pm$0.9 & 6.6$\pm$1.2
& \nodata & 37.1$\pm$2.6 \phn \\

D.3
& 2762.1$~\chi^2$=34.6 & -340 & 41 & 1.0
& 312.0$\pm$8.2 & \nodata & \nodata
& 4.3$\pm$1.0 & 6.3$\pm$1.2
& \nodata & 35.3$\pm$2.8 \phn \\

D.4
& 1886.6$~\chi^2$=19.5 & -317 & 39 & 0.92
& 142.2$\pm$5.5 & \nodata & \nodata
& \nodata & \nodata
& \nodata & 16.2$\pm$1.9 \phn \\

\enddata

\tablecomments{\\
I : Intensity (10$^8$ photons/cm$^2$ sec sr)\\
               V$_{LOS}$, V$_{1/e}$ : km / sec\\
               T$_k$ : 10$^5$K ~~ \\
{$^*$} : Two Gaussian fit\\
$^a$ Observations at the edge of the observed wavelength range. Lower limit of the C III emission.\\
$^b$ FWHM is within the instrumental width. Possibilities of $^a$\\
$^c$ possible to be part of the Ly$\alpha$ continuum. }

\end{deluxetable}

\begin{deluxetable}{lccccccccccc}
\tabletypesize{\scriptsize}
\setlength{\tabcolsep}{0.02in}
\tablecaption{Energy budgets for the three heating models.}
\tablewidth{0pt}
\tablehead{
\colhead{Blob No.} & \multicolumn{7}{c}{O VI (H1)} & \multicolumn{4}{c}{Ly$\alpha$, C III, CII, N III (H2)} \\

\colhead{} & \multicolumn{2}{c}{Q$\propto$Q$_{AHH}$} & \multicolumn{2}{c}{Q$\propto$n} & \multicolumn{3}{c}{Q$\propto$Q$_{KR}$} & \multicolumn{2}{c}{Q$\propto$Q$_{AHH}$} & \multicolumn{2}{c}{Q$\propto$n} \\

\colhead{} & \colhead{H.E.} & \colhead{T.E.}  & \colhead{H.E.} & \colhead{T.E.}  & \colhead{H.E.} & \colhead{T.E.} & \colhead{h} & \colhead{H.E.} & \colhead{T.E.}  & \colhead{H.E.} & \colhead{T.E.}
}

\startdata
A& 4.4-444 & 0.1-9.7 & 5.1-51 & 0.1-0.9 & 87-106 & 0.5 & 81-84
& 11 & 0.3 & 8.0 $^d$ & 0.1 $^d$ \\

B.1& 129-420 & 3.0-9.4 & 57-90 & 1.0-1.6 & 71-72 & 0.3 & 77
& 13-21$^a$ & 0.3-0.5 $^a$ & 14-27$^a$ & 0.3-0.4$^a$ \\

B.2& 58-65 & 1.3 & 57 & 1.0 & 71-77 & 0.3 & 77-78
& 5.5-5.8 & 0.1 & 9.0 & 0.2 \\

B.3& 15-89 & 0.3-2.0 & 9.0-57 & 0.2-1.0 & 71-77 & 0.3 & 77-78
& 5.2-5.5 $^{bd}$ & 0.1$^{bd}$ & 9.0 $^e$ & 0.2 $^e$ \\

C.1& 18-225 & 0.4-5.0 & 25-100 & 0.4-1.7 & 57-102 & 0.2-0.4 & 71-82
& 2.5-45 & 0.07-1.0 & 4.0-25 & 0.07-0.4 \\

C.2& 4.4-44 & 0.1-1.0 & 6.3-40 & 0.1-0.7 & 57-70 & 0.2-0.3 & 71-75
& 3.7-4.1 & 0.1 & 6.3 & 0.1 \\

C.3& 60-68 & 1.4 & 63 & 1.1 & 62-70 & 0.2-0.3 & 73-75
& 3.6-4.3 & 0.1 & 6.3 & 0.1 \\

C.4& 74 & 1.6 & 63 & 1.1 & 70$^c$ & 0.3$^c$ & 75$^c$
& 4.2-4.7$^d$ & 0.1$^d$ & 6.3$^d$ & 0.1$^d$ \\

C.5& 138-144 & 3.4 & 63 & 1.1 & 57-82 & 0.2-0.3 & 71-78
& 5.4-5.7 & 0.1 & 6.3 & 0.1  \\

C.6& 18-45 & 0.3-0.8 & 16-40 & 0.3-0.7 & 57-82 & 0.2-0.3 & 71-78
& 4.0-4.5$^d$ & 0.09$^d$ & 4.0$^d$ & 0.07$^d$ \\

D.1& 51-56 & 1.1 & 17-44 & 0.3-0.7 & 58-71 & 0.2 & 70-74
& 4.9-5.3 & 0.1 & 6.9 & 0.1 \\

D.2& 9.0-57 & 0.2-1.0 & 6.9-44 & 0.1-0.7 & 58-105 & 0.2-0.3 & 70-81
& 4.7-5.2$^{de}$ & 0.1$^{de}$ & 6.9$^{de}$ & 0.1$^{de}$ \\

D.3& 6.7-65 & 0.1-1.3 & 6.9-44 & 0.1-0.7 & 58-59 & 0.2 & 70
& 6.2-6.3$^d$ & 0.1$^d$ & 6.9$^d$ & 0.1$^d$ \\

D.4& 5.1-32 & 0.09-0.5 & 4.4-27 & 0.07-0.5 & 58-218 $^c$ & 0.2-0.7 $^c$ & 70-90$^c$
& 4.0-4.6$^e$ & 0.09$^e$ & 4.4 & 0.07 \\

\enddata
\tablecomments{\\
T.E. and H.E. : Thermal Energy and Heating Energy (1.$\times$10$^{14}$ erg/g) assuming 10$\%$ Helium \\
h : the percentage of the lost magnetic energy appearing as heat \\
$^a$ : Does not satisfy the C II observations \\
$^b$ : Does not satisfy the N III observations \\
$^c$ : Does not satisfy the O VI / O V constraint \\
$^d$ (and $^e$) : Does not satisfy the C II $<$ 3.* C II (and C III) criterion \\
}

\end{deluxetable}

\begin{figure}
\epsscale{0.8} \plotone{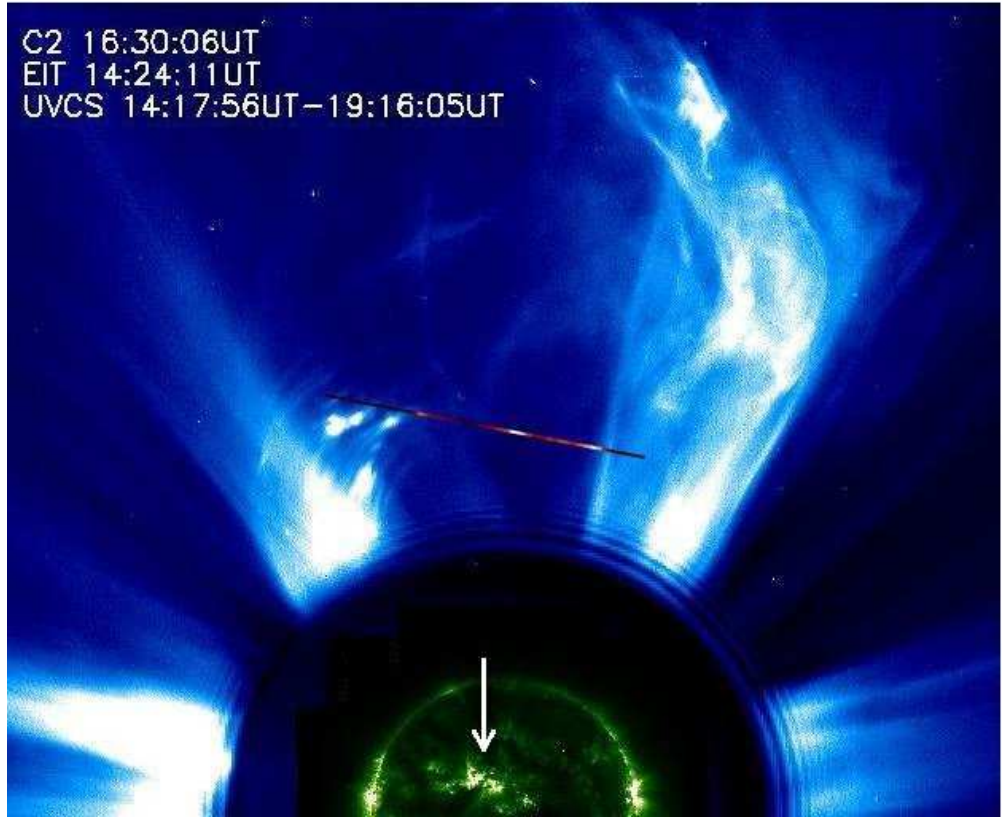}
\caption{Composite image of EIT, UVCS and LASCO C2. The UVCS slit image shows the intensity distribution of the O VI line summed over exposures from 14:17:56UT to 19:16:05UT at 2.4 R$_\sun$ and P.A.=349$^\circ$. 
The arrow indicates the active region where the bright flare was observed. 
\label{fig 1}}
\end{figure}

\begin{figure}
\epsscale{0.5} \plotone{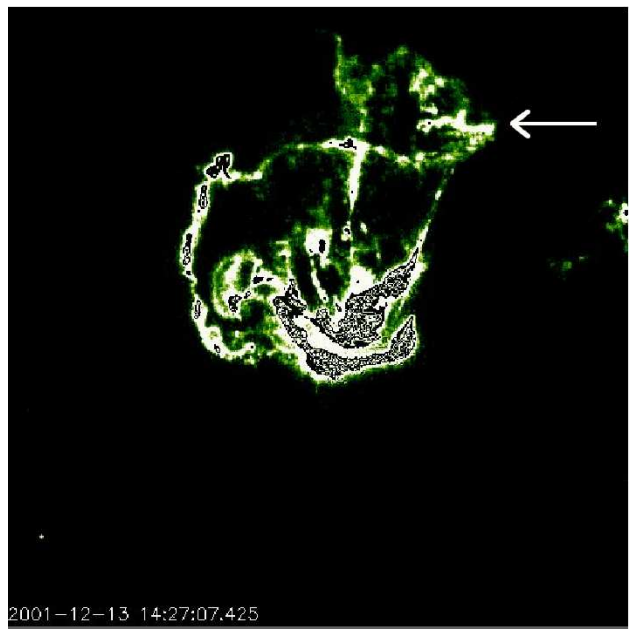}
\caption{TRACE 1600 \AA\/ at 14:27:07 UT. The arrow indicates the filament eruption. The lower part of the active region was seen as dark due to a saturation. \label{fig 2}}
\end{figure}

\begin{figure}
\epsscale{0.8} \plotone{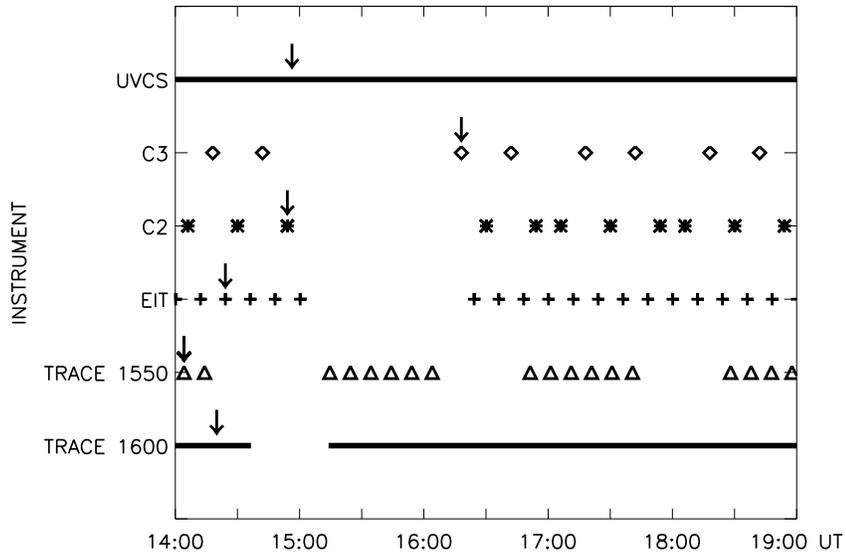}
\caption{Observation time of each instrument. The arrows represent the first detection of this event at each instrument. In the TRACE 1550 \AA\ observation, the bright emission was already started at 14:04 UT. \label{fig 3}}
\end{figure}

\begin{figure}
\epsscale{0.8} \plotone{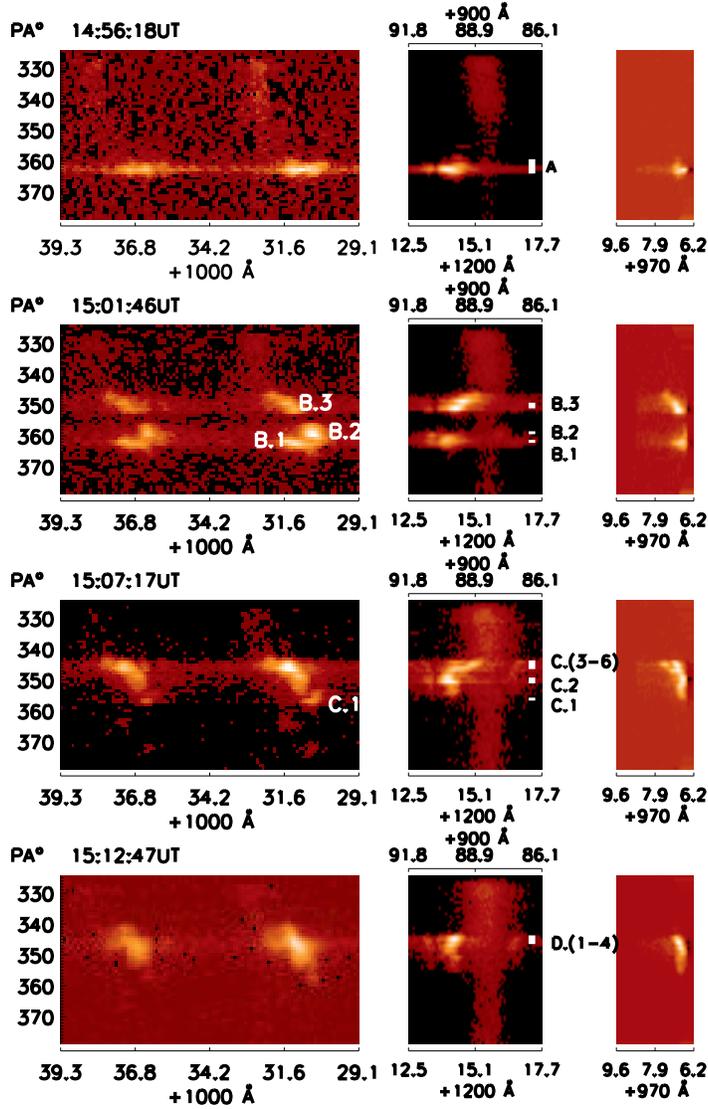}
\caption{UVCS observations of the December 13 2001 CME. The left panel shows the O VI 1032 \AA\ and 1037 \AA\/. The middle panel shows the Ly$\alpha$, [O V], O V], N III lines in both primary and redundant wavelengths. The primary (991.8 \AA\ - 986.1 \AA\ ) and redundant wavelengths (1212.5 \AA\ - 1217.7 \AA\ ) are represented in the upper and bottom axes in the middle panel, respectively. The right panel shows the C III line. .\label{fig 4}}
\end{figure}

\begin{figure}
\epsscale{1.0} \plotone{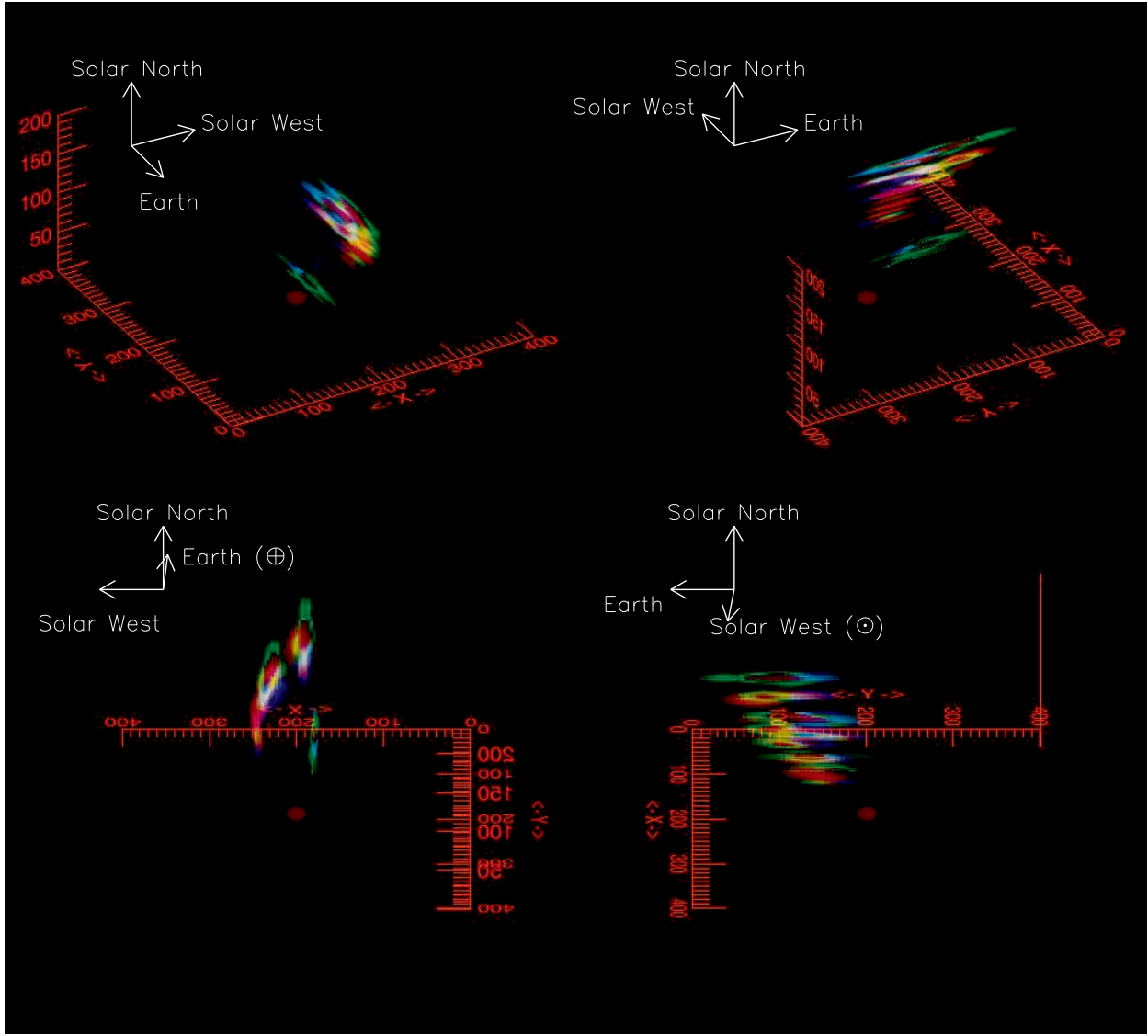}
\caption{Three-dimensional distribution of the O VI (red), Ly$\alpha$ (green), and C III (blue) emission. Each grid presents 0.1 solar radius and the Sun is placed at the coordinate, (x:200, y:200, z:0). This figure is also available as an mpeg animation. The movie consists of two rotations, the first is the expansion of CME from 14:56 UT to 19:16 UT and the second rotation is at 19:16 UT. \label{fig 5}}
\end{figure}

\begin{figure}
\epsscale{1.0}\plotone{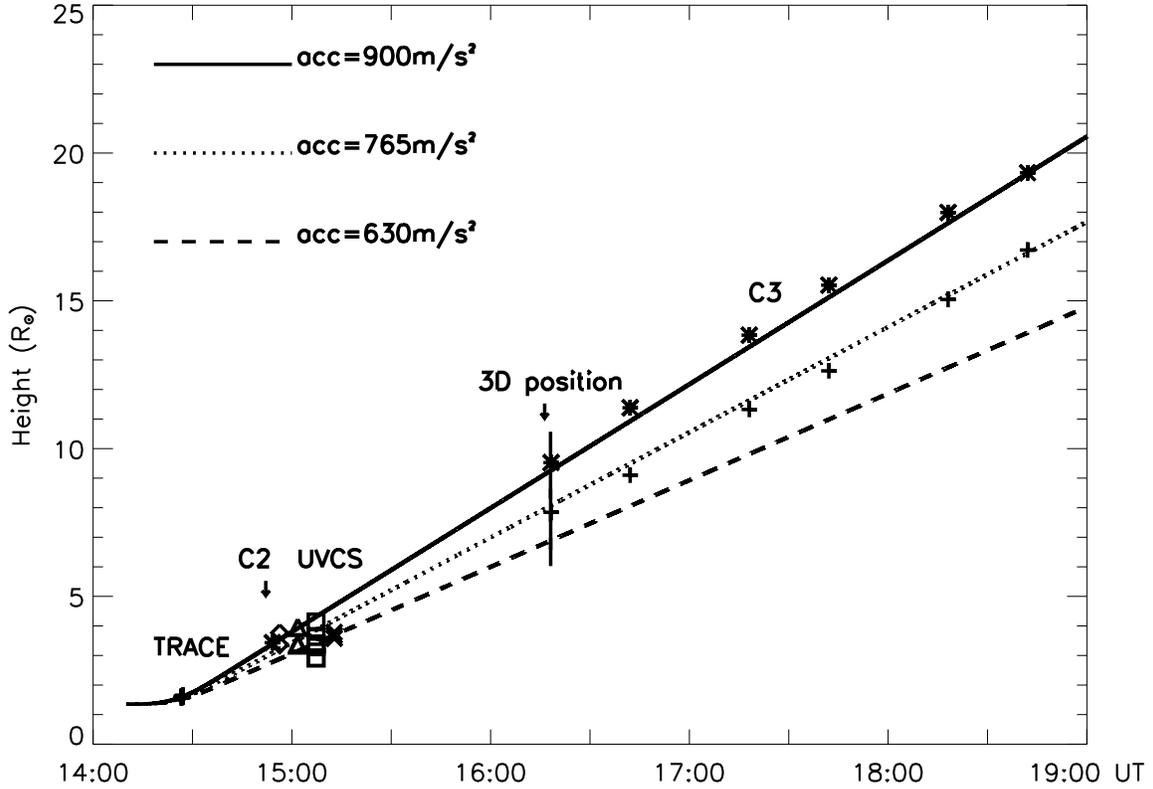}
\caption{Expansion heights of the CME plasma. The stars and crosses for C3 represent the position of the upper edge of left loop (where blob 'A' is placed) and right loop(where blob 'C' is placed) in the LASCO image in Figure 9. For the UVCS observation, the diamond represents blob 'A', the triangle represents blob 'B', the square represents blob 'C', and the mark, 'x', represents for blob 'D'. The solid line is the distance applied for blob A, the dotted line is the distance applied for blob B, and the dashed line is the distance for blobs C and D.\label{fig 6}}
\end{figure}

\begin{figure}
\epsscale{0.5} \plotone{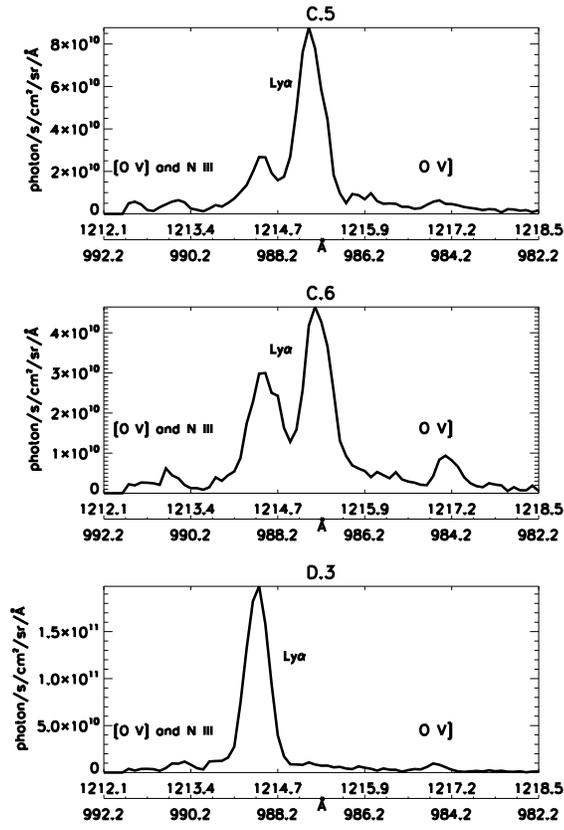}
\caption{Ly$\alpha$, [O V], O V], and N III line intensities of the blobs, C.5, C.6, and D.3.\label{fig7}}
\end{figure}

\begin{figure}
\epsscale{1.0} \plotone{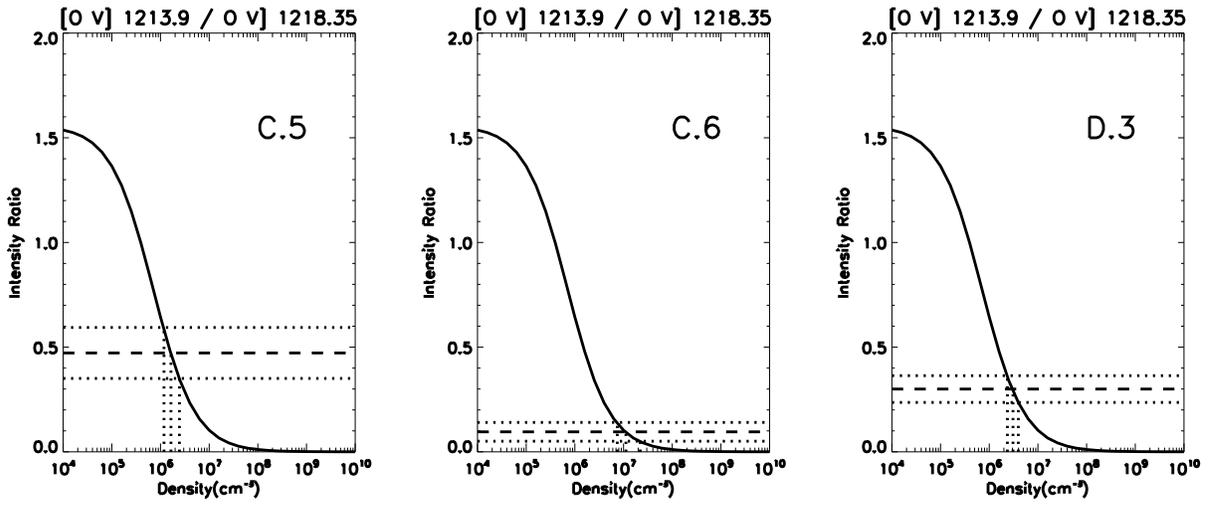}
\caption{The observed line ratio of the [O V]/O V] and the electron number density evaluated from CHIANTI 5.2.\label{fig 8}}
\end{figure}

\begin{figure}
\plotone{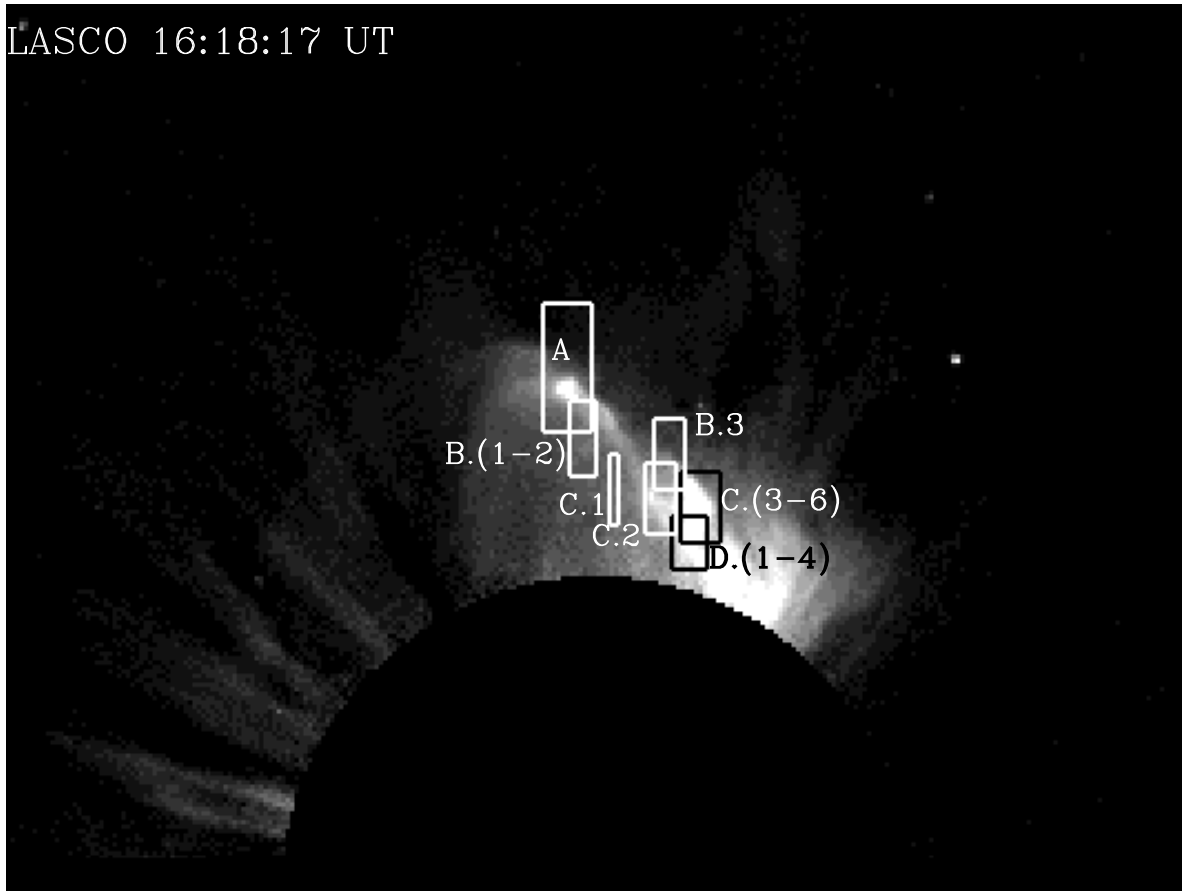}
\caption{The reconstructed positions of the 14 blobs are presented in the LASCO C3 image.\label{fig 9}}
\end{figure}

\begin{figure}
\epsscale{1.1} \plotone{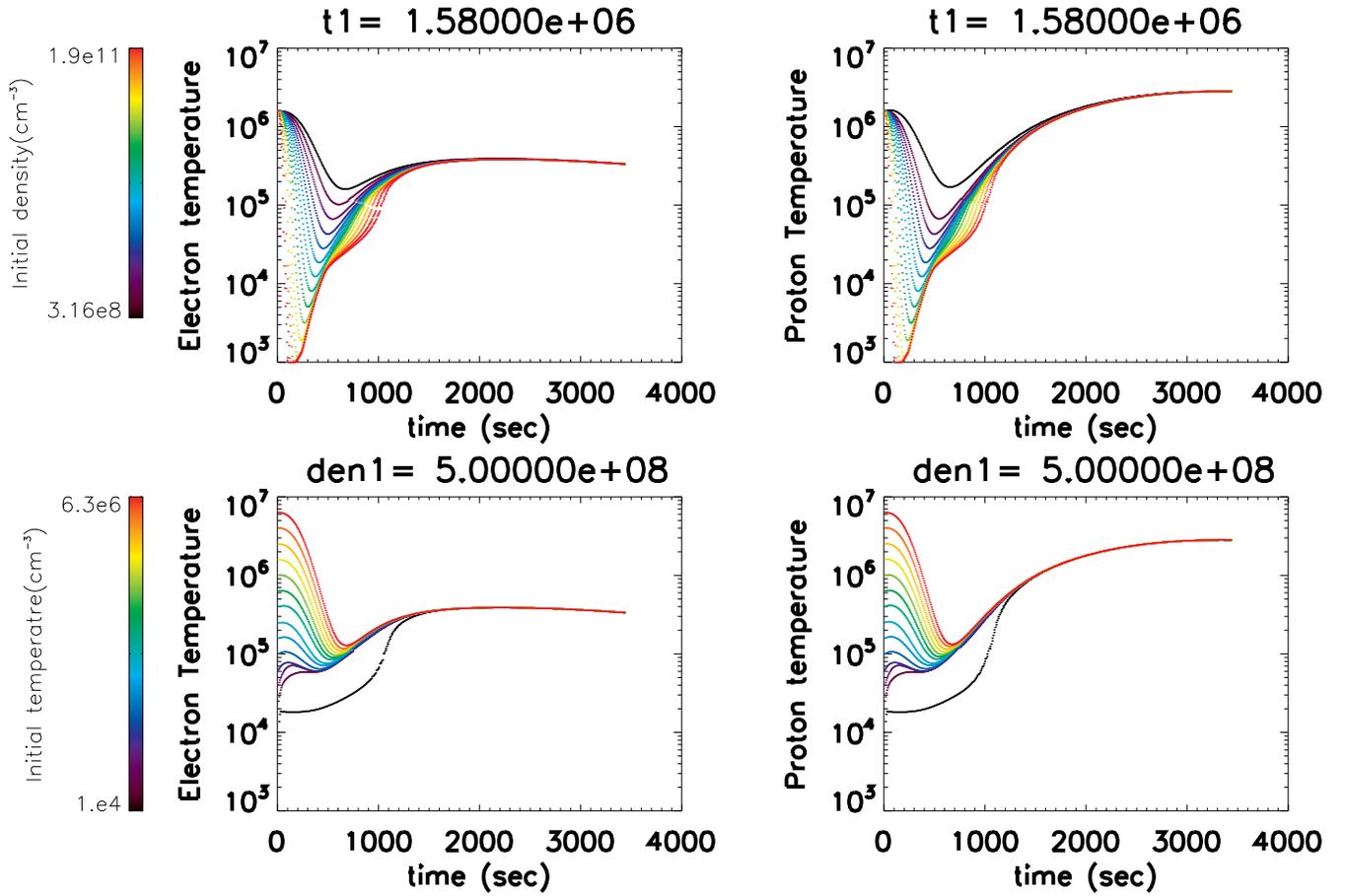}
\caption{Temperature evolution of the protons and electrons of the model Q$_{AHH}$ for matching the O VI and O V] line intensities of the blob C.5. \label{fig 10}}
\end{figure}

\begin{figure}
\epsscale{1.1} \plotone{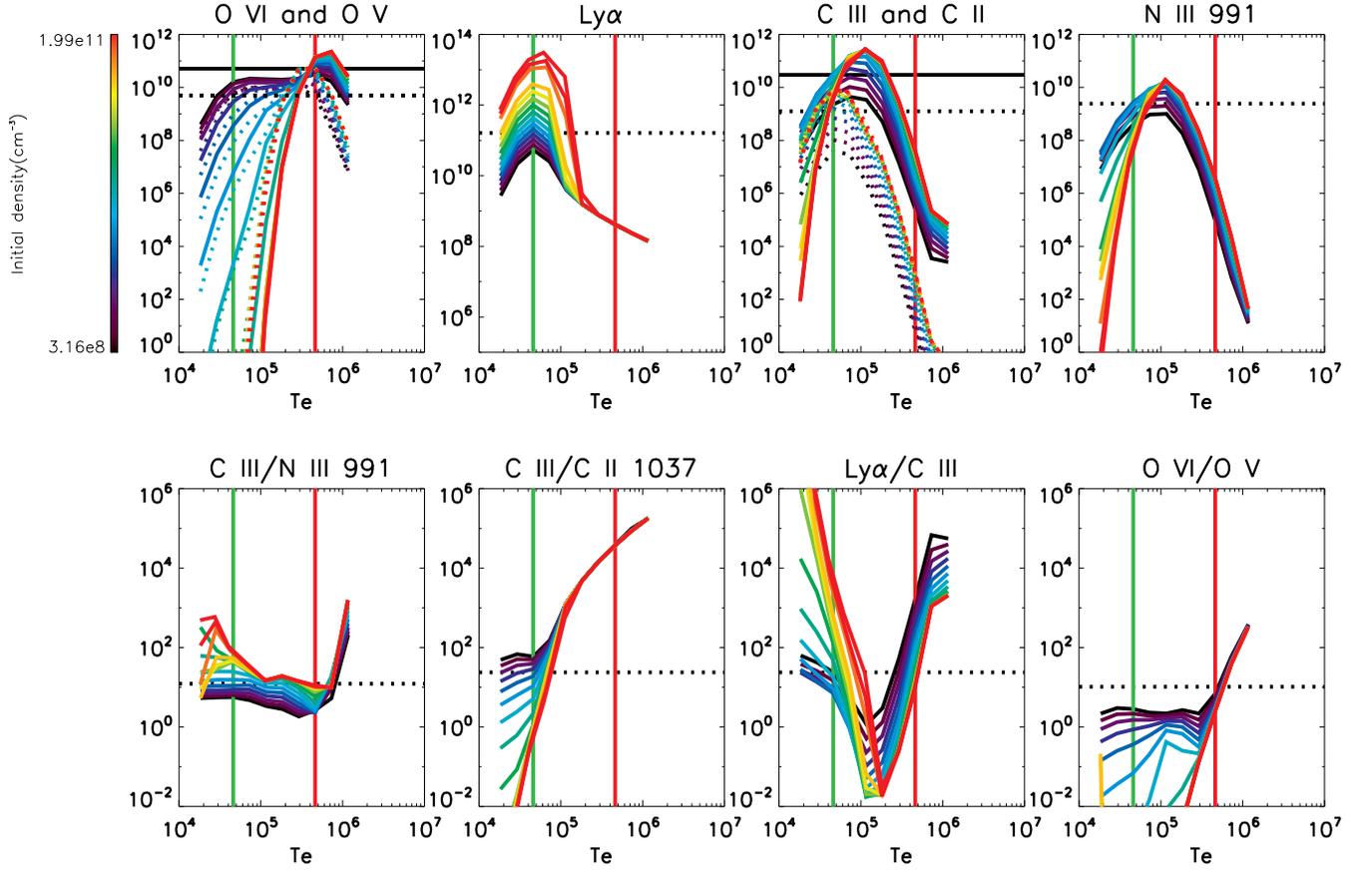}
\caption{The line intensities when different heating rates are applied to the heating model, Q$\propto$Q$_{AHH}$. The rainbow colors indicate different densities. The horizontal lines are the observed values. In the first panel, the solid and dotted lines are the observed O VI and O V values, respectively. In the upper third panel, the solid and dotted lines are for the C III and C II lines, respectively. The green vertical line is the temperature that match the Ly$\alpha$, C III, C II, and N III emissions and the red vertical line is the temperature that match the O VI/ O V and O VI emissions.\label{fig11}}
\end{figure}

\begin{figure}
\epsscale{1.0} \plotone{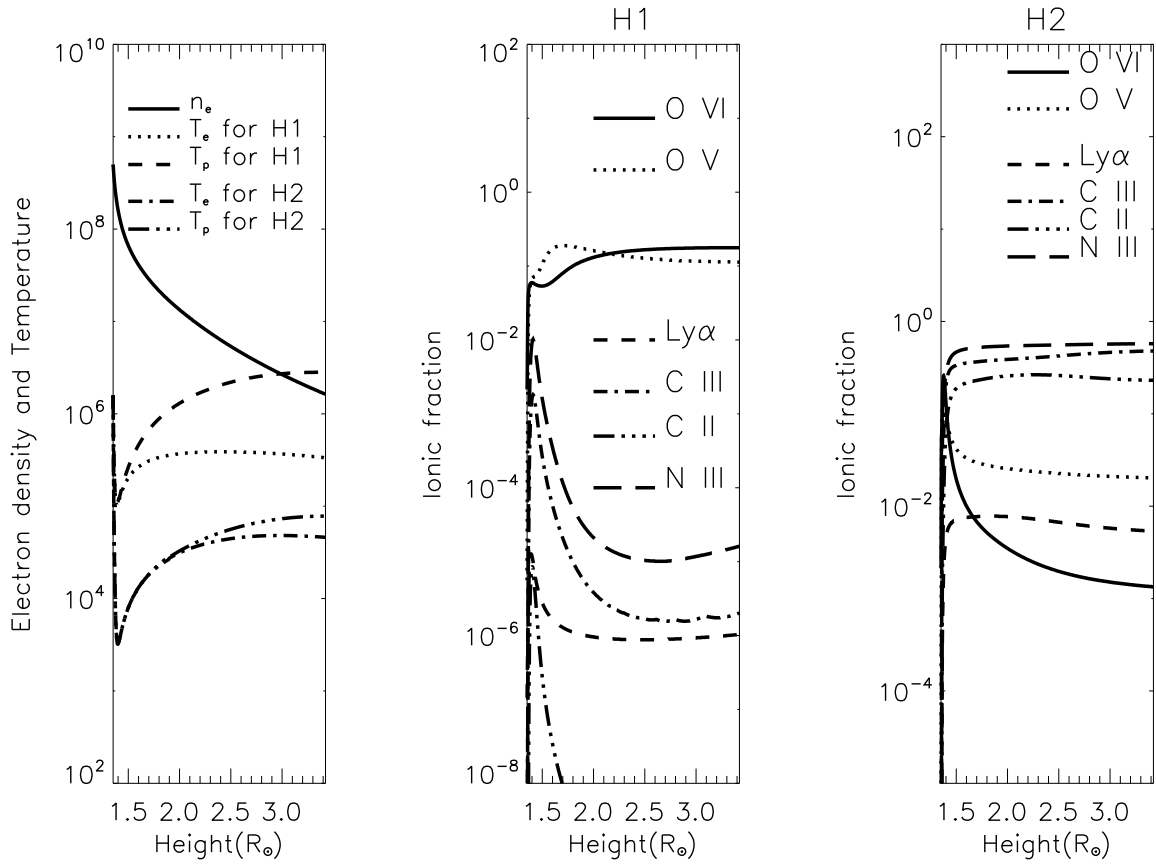}
\caption{The evolution of the density, temperature, and ionization states for one of the acceptable model for blob C.5.\label{fig 12}}
\end{figure}

\begin{figure}
\epsscale{1.0} \plotone{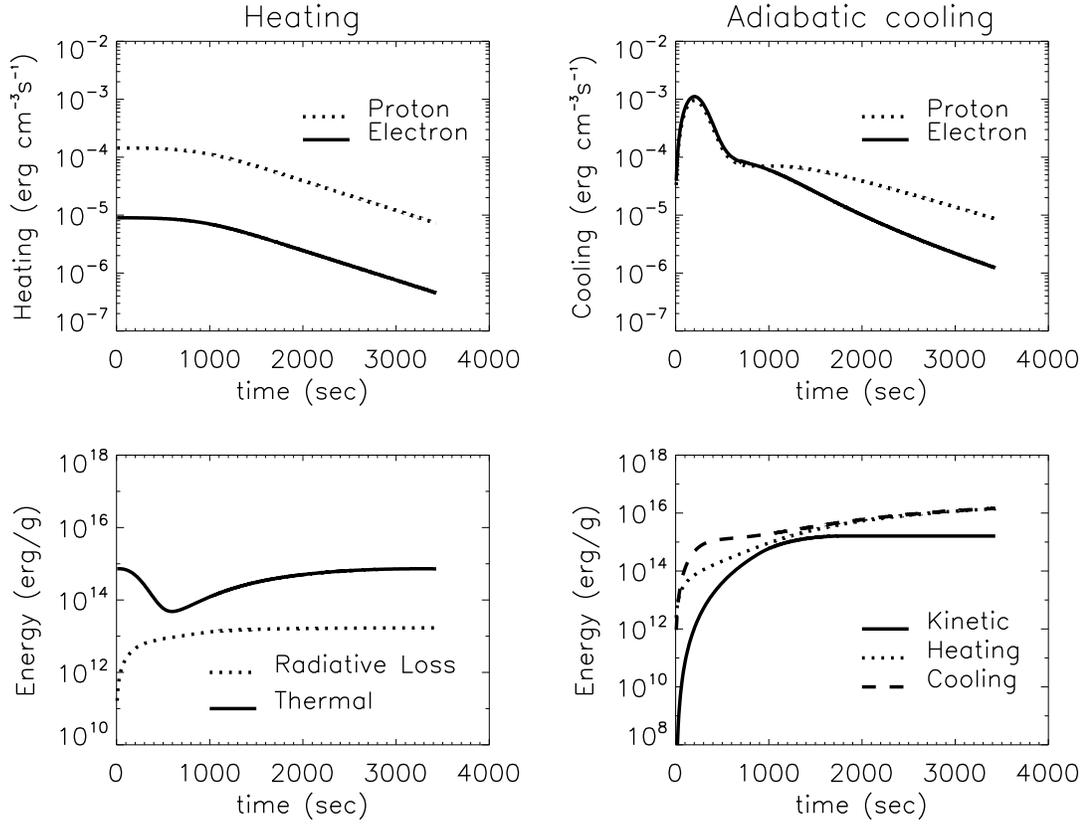}
\caption{Energy budgets for H1, Q$\propto$Q$_{AHH}$.\label{fig 13}}
\end{figure}

\begin{figure}
\epsscale{1.0} \plotone{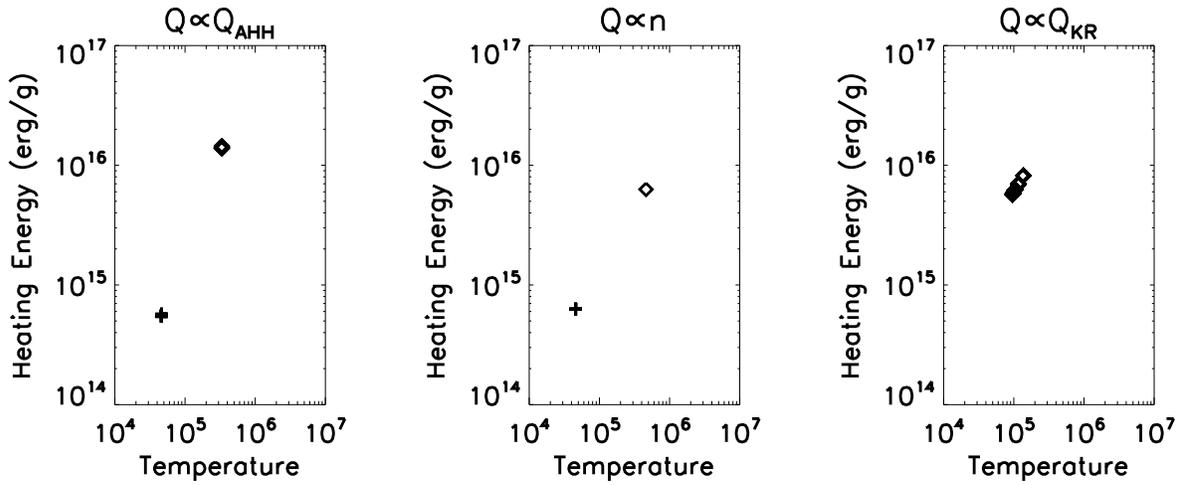}
\caption{Heating Energy. $\diamond$ and $+$ are for H1 and H2, respectively.\label{fig 14}}
\end{figure}

\clearpage

\begin{figure}
\epsscale{1.0} \plotone{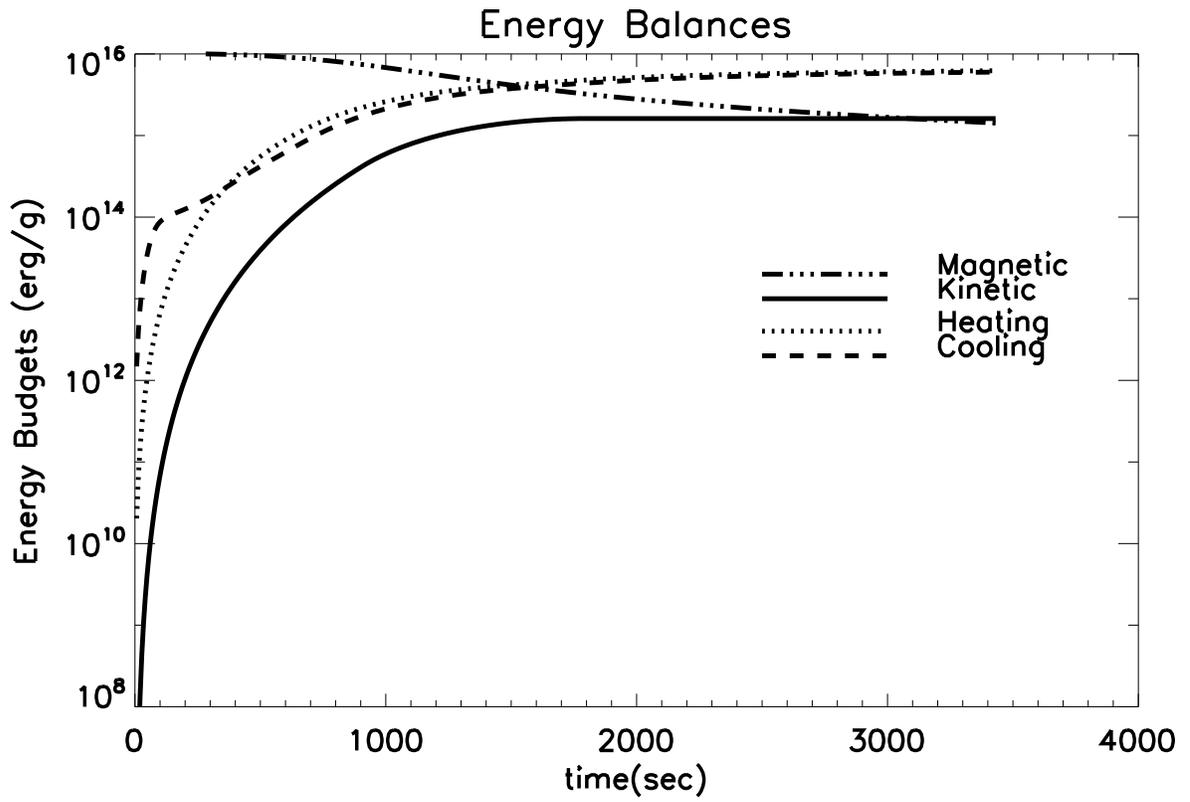}
\caption{Energy balance of the Kumar and Rust's heating model.\label{fig 15}}
\end{figure}

\begin{figure}
\epsscale{1.0} \plotone{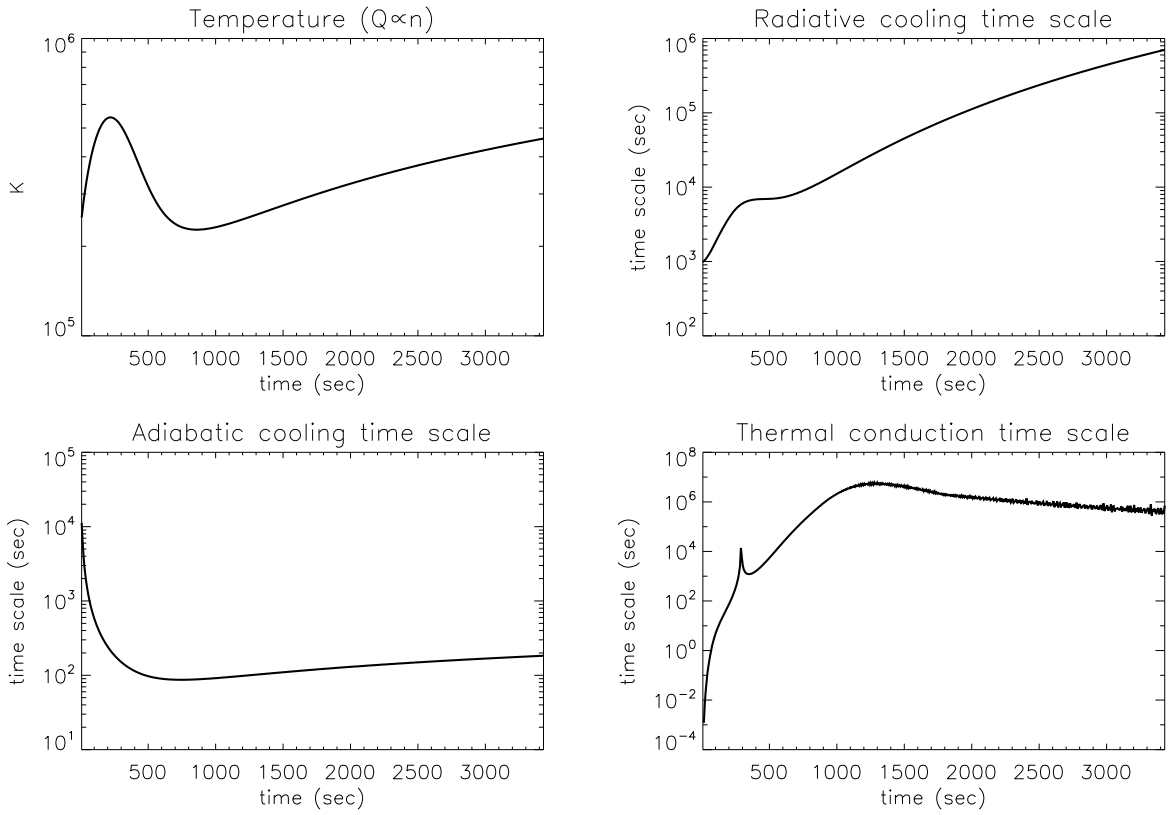}
\caption{Time scale of radiative cooling, adiabatic cooling and thermal conduction.\label{fig 16}}
\end{figure}

\clearpage

\end{document}